\documentclass[journal]{IEEEtran}
\usepackage{cite}

\ifCLASSINFOpdf
   \usepackage[pdftex]{graphicx}
\else
   \usepackage[dvips]{graphicx}
\fi
\usepackage{amsmath}
\ifCLASSOPTIONcompsoc
  \usepackage[caption=false,font=normalsize,labelfont=sf,textfont=sf]{subfig}
\else
  \usepackage[caption=false,font=footnotesize]{subfig}
\fi
\usepackage{url}

\usepackage{stfloats}

\begin{document}
%
\title{Thermal Characterization of Tl$_2$LiYCl$_6$:Ce (TLYC)}
%
%
%

\author{Maya~M.~Watts,
        Katherine~E.~Mesick,
        Kurtis~D.~Bartlett,
        and~Daniel~D.S.~Coupland
\thanks{Manuscript received XX; revised XX.}
\thanks{Research presented in this paper was supported by the DOE Office of Science Summer Undergraduate Laboratory Internship (SULI) program, the Laboratory Directed Research and Development program of Los Alamos National Laboratory under project number 20170438ER, and the US Department of Energy through the Los Alamos National Laboratory. Los Alamos National Laboratory is operated by Triad National Security, LLC, for the National Nuclear Security Administration of U.S. Department of Energy (Contract No. 89233218CNA000001).}
\thanks{The authors are with the Space Science and Applications group at Los Alamos National Laboratory, Los Alamos, NM 87545 USA.  (Corresponding author email: kmesick@lanl.gov)}
}

%
%

\markboth{Journal of \LaTeX\ Class Files,~Vol.~14, No.~8, August~2015}%
{Shell \MakeLowercase{\textit{et al.}}: Bare Demo of IEEEtran.cls for IEEE Journals}
%



\maketitle

\begin{abstract}
Tl$_2$LiYCl$_6$:Ce (TLYC) is a new dual-detection elpasolite scintillator that can detect and distinguish between gamma rays and neutrons using pulse-shape discrimination (PSD).  It has a higher density and Z-number than the more mature and well-known elpasolite Cs$_2$LiYCl$_6$:Ce (CLYC), causing it to have a significantly better gamma-ray stopping power.  These properties make TLYC an attractive alternative to CLYC for resource-constrained applications where size and weight are important, such as space or national security applications.  Such applications may be subjected to a wide range of temperatures, and therefore TLYC's performance was characterized for the first time over a temperature range of $-$20$^{\circ}$C to $+$50$^{\circ}$C in 10$^{\circ}$C increments.  TLYC's thermal response effects on light-output linearity with energy, gamma-ray photopeak energy resolution, detected neutron energy, pulse shapes, and figure of merit is analyzed and reported.  The light output of TLYC was found to be linear with energy over the tested temperature range and was observed to decrease with increasing temperature.  The decay time of the scintillation light output was observed to decrease with decreasing temperature at short times, leading to a decreasing PSD figure of merit.  The gamma-ray photopeak energy resolution was also observed to degrade with decreasing temperature, due to an asymmetric broadening of the photopeak at low temperatures.
\end{abstract}


%
\IEEEpeerreviewmaketitle

\section{Introduction}
%
%
%
%
\IEEEPARstart{S}{everal} scintillators from the elpasolite class of crystals are exciting candidates for creating the next generation of low-resource particle detectors for space and national security applications. Depending on the specific material, they have moderate to high light output (20,000 - 60,000 photons/MeV \cite{Glodo2011}) and good energy resolution (as good as 2.9\% at 662 keV \cite{Glodo2011}), which provides good signal-to-noise and peak detection ability. Certain elpasolite scintillators have specifically garnered attention for their sensitivity to both gamma rays and neutrons and the ability to distinguish between incident particles species using pulse-shape discrimination (PSD). Their neutron sensitivity arises from constituent materials that have high thermal neutron absorption cross sections or high cross sections for fast neutron (n,p) reactions, an example of which would be elpasolites containing $^{6}$Li that undergo the $^6$Li(n,$\alpha$)T reaction with a Q value of 4.8 MeV.  The dual-detection elpasolite scintillators are of specific interest for applications where size and mass are at a premium, such as hand-held detectors for radioisotope detection and radiation detection in space.  Both of these applications require sensor operation over a wide range of temperatures.

There are a variety of different dual-detection elpasolite scintillators, the most well-known of which is Cs$_2$LiYCl$_6$:Ce (CLYC).  Compared to traditional scintillators such as CsI(Tl), CLYC offers better gamma-ray energy resolution (as good as 3.9\% at 662~keV \cite{Glodo2011}) and the ability to detect neutrons with a superb PSD figure of merit (FOM) $>$4 \cite{Lee2012}, where a larger FOM indicates more PSD separation between neutrons and gamma rays.  However, CLYC suffers from lower gamma-ray photopeak detection efficiency due to a lower effective atomic number of the material, $Z_{eff}$ = 45 for CLYC versus $Z_{eff}$ = 54 for CsI.  The lower $Z_{eff}$ means that larger volumes are required to achieve the same gamma-ray detection efficiency, partially negating the size advantage enabled by the dual-detection capability.  Many other elpasolite scintillators have been investigated and a few have recently been commercialized in part to improve the gamma-ray detection efficiency while retaining or improving the energy resolution and maintaining adequate FOM.  Examples are the related Cs$_2$LiLaBr$_6$:Ce (CLLB) and Cs$_2$LiLa(Br,Cl)$_6$:Ce (CLLBC) materials which currently offer the best energy resolution available from the dual-detection elpasolite scintillators at a somewhat improved $Z_{eff}$ = 47, but suffer from the intrinsic $\alpha$-particle background common to lanthanum-containing scintillators \cite{Woolf2016,Mesick2017,Hull2019}. 

Tl$_2$LiYCl$_6$:Ce (TLYC) is an emerging dual-detection elpasolite scintillator with quite high $Z_{eff}$ = 69 and no lanthanum component.  These properties make TLYC an attractive alternative to CLYC, CLLB, and CLLBC when the driving factors are gamma-ray detection efficiency, size, and mass.  Relative to CLYC, the mean penetration depth of 662~keV gamma rays in TLYC is $\sim$40\% shorter \cite{Hawrami2016}, and the 662~keV photopeak efficiency $\sim$4.4$\times$ and 3.2$\times$ higher for 1~cm and 1'' thick crystals, respectively.  Previous measurements of TLYC have reported an energy resolution as good as 3.8\% full width half maximum (FWHM) at 662~keV and a good FOM of $\sim$2 with a light output of $\sim$25,000 - 29,000~ph/MeV \cite{Hawrami2016,Hawrami2017}.  However, the performance of TLYC over a large range of temperatures, such as required for national security and space applications, has not yet been studied.  That study is the focus of the current work.

\section{Experimental Methods}

To test the sensitivity of TLYC's performance to temperature, a TLYC crystal was coupled with a photomultiplier tube (PMT) and subjected to temperatures between $-20^{\circ}$C to $+50^{\circ}$C in 10$^{\circ}$C steps. This temperature range was chosen to satisfy the ANSI standard for Hand-Held Instruments for the Detection and Identification of Radionuclides\cite{ANSI_N4234_2015}; it is also a reasonable temperature range for some space missions. The light-output linearity with energy, gamma-ray photopeak energy resolution, detected neutron energy, pulse shapes, and figure of merit were measured at each temperature to assess performance.

\begin{figure}[t]
\centering
\includegraphics[width=3in]{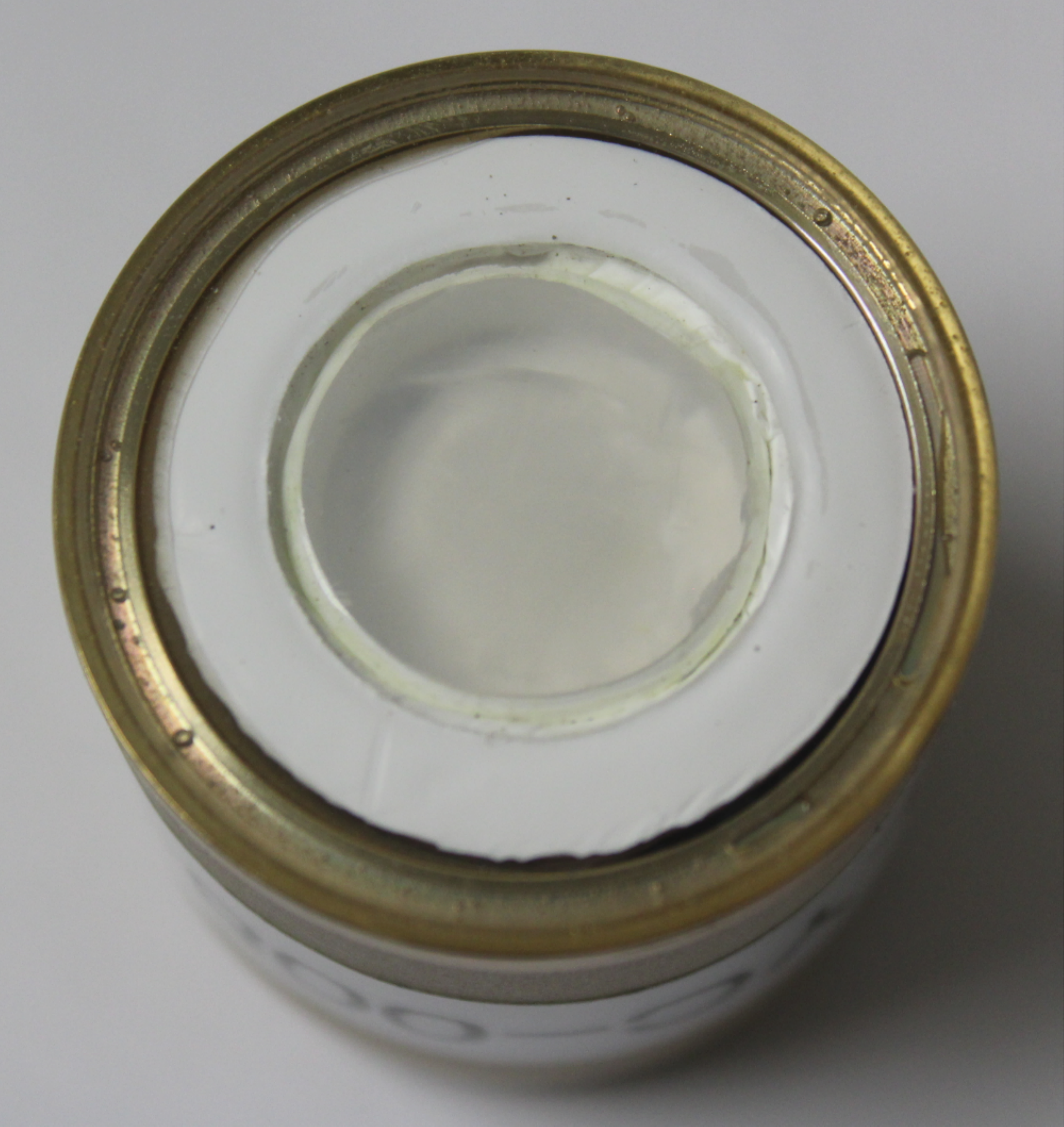}
\caption{Picture of the packaged TLYC crystal used in this study.}
\label{fig:crystal}
\end{figure}

A 3'' super-bialkali Hamamatsu 6233-100 PMT was coupled with a TLYC crystal obtained from Radiation Monitoring Devices in hermetically sealed aluminum packaging with a quartz window, shown in Fig.~\ref{fig:crystal}.  The crystal had a diameter of 15~mm and a length of 11~mm.  Visually, the TLYC crystal is fairly transparent with a small crack towards the bottom of the sample relative to the window (seen in Fig.~\ref{fig:crystal} towards the top).

\begin{figure}[h]
\centering
\includegraphics[width=3.5in]{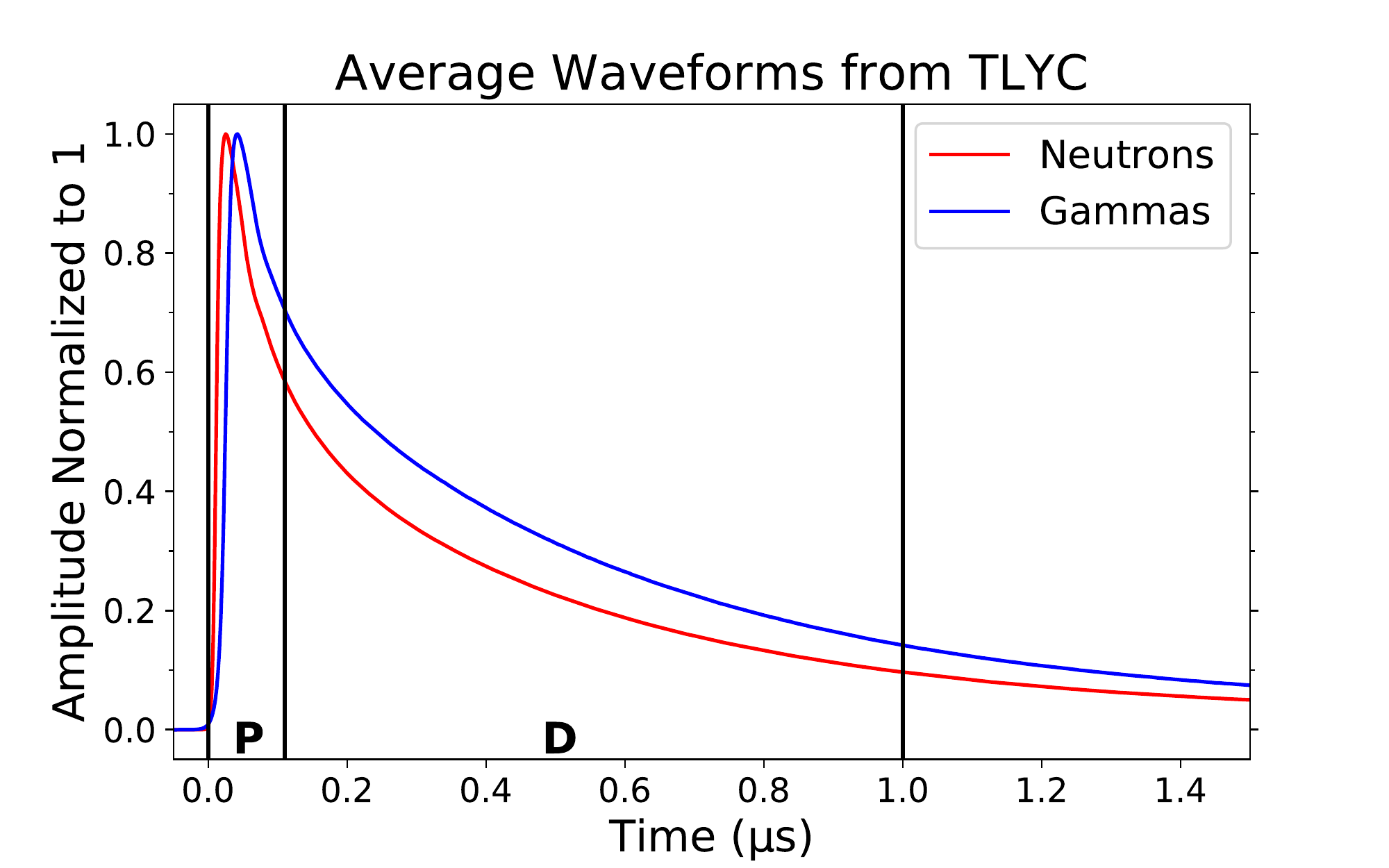}
\caption{The light-output time profile for neutrons (red) and gammas (blue) measured from TLYC during this study at 20$^{\circ}$C with example prompt (P) and delayed (D) integration windows indicated.}
\label{fig:light}
\end{figure}

The interaction of gamma rays and neutrons with TLYC results in different combinations of scintillation mechanisms, which causes the time profile of the light output to differ based on incident particle species as shown in Fig.~\ref{fig:light}. The incident particle can then be identified through pulse-shape discrimination.  In TLYC, PSD can be performed by assigning prompt and delayed integration windows to regions of the pulse and a PSD ratio is calculated by dividing the prompt integration (P) by the delayed integration (D), a common charge-integration method that has been used for other dual-detection elpasolite scintillators \cite{Glodo2011}.  The gammas have a slower pulse decay time than the neutrons, causing the neutrons and gammas to have different PSD ratios and therefore allowing robust particle classification to be achieved.  A PSD figure of merit is defined by taking the difference in the means ($\mu$) divided by the sum of the full width half maximums ($\Gamma$) of Gaussian fits to the neutron ($n$) and gamma ($\gamma$) PSD ratio:
\begin{equation}\label{eq1}
\textrm{FOM} = \frac{\mu_n - \mu_{\gamma}}{\Gamma_n + \Gamma_{\gamma}}~.
\end{equation}

To characterize the light-output linearity with energy, gamma-ray photopeak energy resolution, and detected energy of the neutron capture, $^{22}$Na and $^{137}$Cs gamma-ray sources and a moderated $^{252}$Cf neutron source were used. The spectra were read out by a DT5730 CAEN waveform digitizer operating in list mode.  The spectra were integrated over 5~$\mu$s for all temperatures except $-10^{\circ}$C and $-20^{\circ}$C where 10~$\mu$s was used since 5~$\mu$s was insufficient to capture the full light output at the colder temperatures.  To characterize the pulse shapes and FOM, 50,000 waveforms were collected with an Agilent Acqiris DC282 waveform digitizer sampling at 500 MegaSamples/s using a moderated $^{252}$Cf source. The pulse shapes were acquired over a 10~$\mu$s window.  

\begin{figure}[b]
\centering
\includegraphics[width=3.5in]{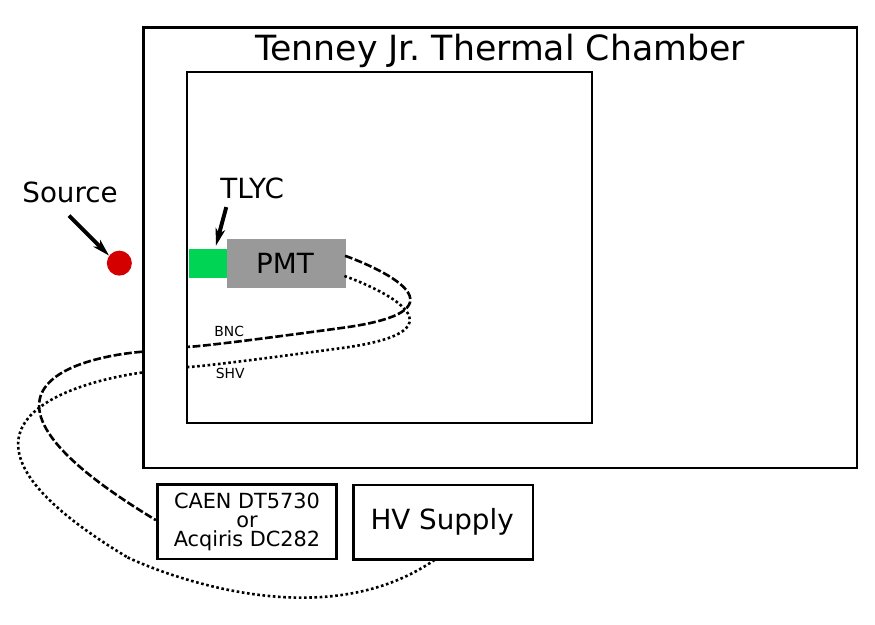}
\caption{Cartoon of the experimental setup.}
\label{fig:setup}
\end{figure}

A schematic of the experimental setup is shown in Fig.~\ref{fig:setup}.  The PMT and packaged TLYC crystal were coupled with optical grease and the PMT biased to $-1425$V.
The packaged TLYC crystal and coupled PMT were placed in a Tenney Jr. thermal chamber.  Cycling between temperatures began at a centroid of $20^{\circ}$C and shifted to either side in $10^{\circ}$C steps (\textit{e.g.}, following a pattern $20^{\circ}$C, $10^{\circ}$C, $30^{\circ}$C, $0^{\circ}$C, and so on) as a precaution against damaging the crystal by starting at one of the temperature extremes due to the unknown thermal resiliency of the crystal. The thermal chamber used a temperature ramp rate of $0.2^{\circ}$C/minute to transition between temperature settings. Once the chamber reached a particular setting, the temperature was held constant for at least six hours (and often overnight) to allow the packaged crystal and PMT to reach thermal equilibrium before data acquisition.  Two methods were used to monitor the temperature of the environment inside of the thermal chamber - 1) a thermocouple readout integrated into the thermal chamber, which had an uncertainty of $\pm 1^{\circ}$C, and 2) an MSR175 data logger located next to the packaged crystal, which carries an accuracy of $\pm0.5^{\circ}$C. Both devices produced readings that agreed within their uncertainties.  The RMD packaging of the crystal was the same as has been used in previous thermal cycling measurements, and there is no expectation of the packaging affecting the thermal performance of the crystal.

\section{Results and Analysis}

\subsection{Spectra features and fitting method}

\begin{figure}[b]
\centering
\includegraphics[width=3.5in]{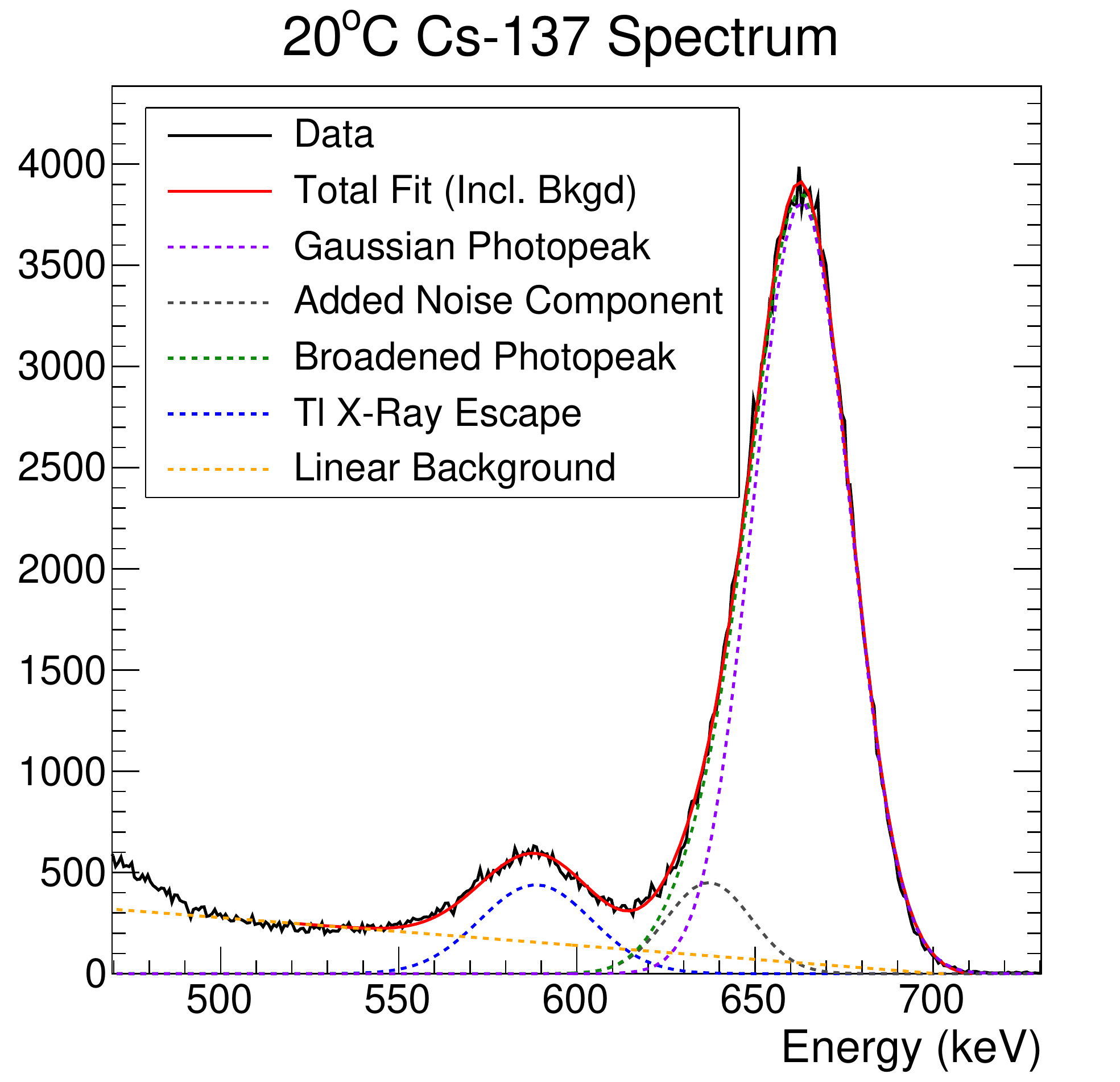}
\caption{Example $^{137}$Cs spectrum at 20$^{\circ}$C showing the triple-Gaussian fit plus linear background and identifying key features in the spectrum.}
\label{fig:spectra}
\end{figure}
Before going into detail on the results of the thermal dependence of TLYC, the features observed in the gamma-ray spectra are described.  Fig.~\ref{fig:spectra} shows the triple-Gaussian fit plus linear background adopted in this work to fit the gamma-ray spectra.  As pointed out in previous publications \cite{Kim2016,Hawrami2016,Hawrami2017}, an X-ray escape peak from the Tl K-edge (85.5~keV) is observed $\sim$74~keV below the photopeak.  This is the first Gaussian of the fit.  The second two Gaussians describe the photopeak.  Near room temperature and above, the gamma-ray spectra can easily be fit with a double-Gaussian fit (photopeak + X-ray escape) and a good fit obtained.  However, as will be discussed more in Section~\ref{sec:eres}, the measured photopeak broadens asymmetrically to lower energy as the temperature is decreased.  We attempted a skew-Gaussian fit of the photopeak, but were unable to constrain the fit at low temperature and obtain a good fit.  Therefore, we describe the overall photopeak (Broadened Photopeak in Fig.~\ref{fig:spectra}) as the sum of two Gaussians - a primary component (Gaussian Photopeak in Fig.~\ref{fig:spectra}) plus an additional component that serves to broaden the photopeak asymmetrically toward low energy (Added Noise Component in Fig.~\ref{fig:spectra}).  It is likely the X-ray escape peak is broadened as well, but due to its lower amplitude it is harder to constrain and therefore fit with a single Gaussian.  The triple-Gaussian fit is constrained in two ways: 1) the mean of the X-ray escape peak was fixed to be 74.6~keV below the primary photopeak (this constraint was determined at $50^{\circ}$C, where the photopeak broadening is negligible) and 2) the width of the X-ray escape peak was constrained to be within $\pm$15\% of the primary photopeak width, on the basis that the primary photopeak and escape peak energy resolution should be similar based on $E^{-1/2}$ scaling.

\subsection{Linearity}\label{sec:lin}

\begin{figure}[h]
\centering
\includegraphics[width=3.5in]{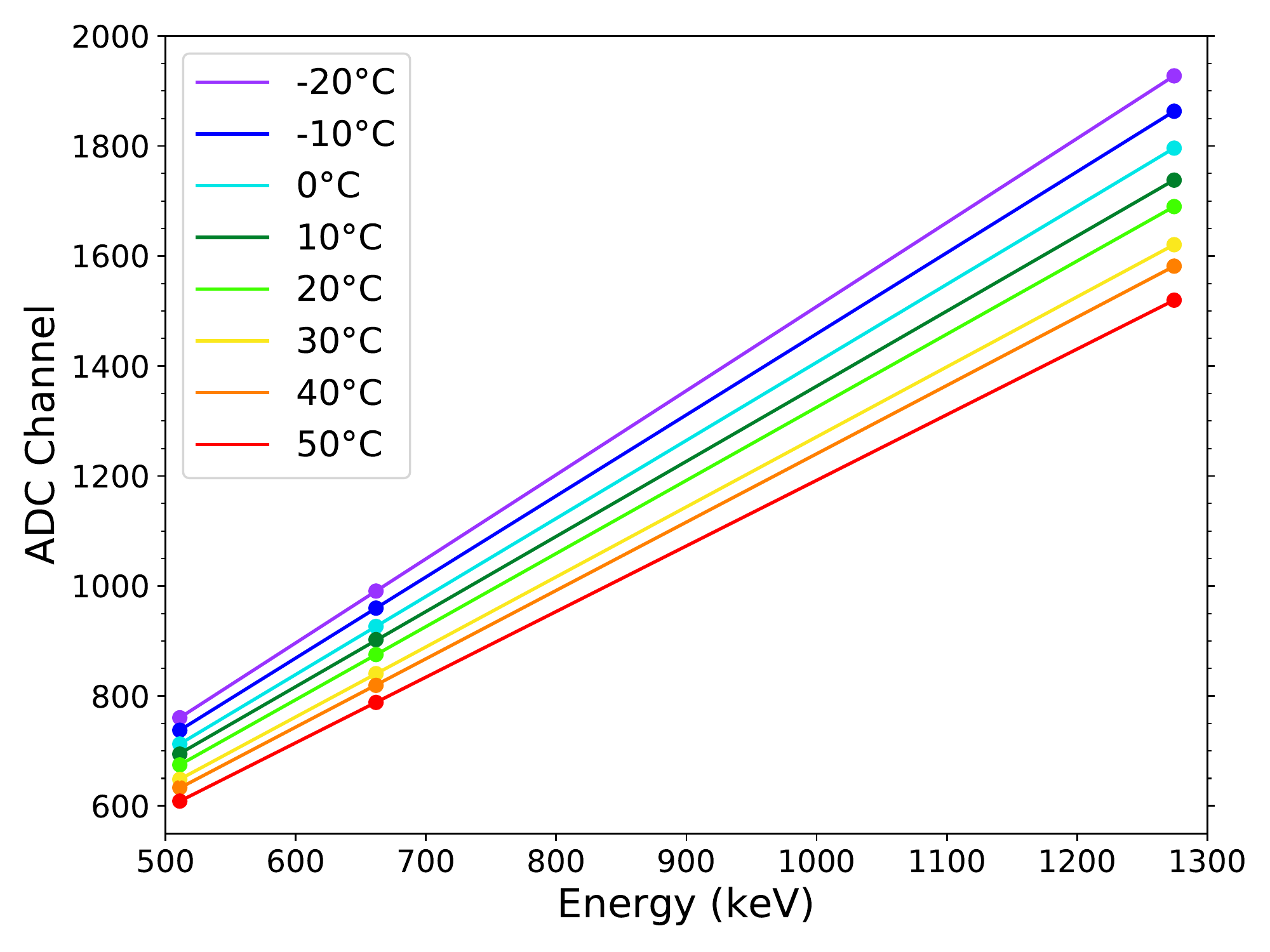}
\caption{The mean ADC channel of the $^{22}$Na and $^{137}$Cs photopeaks as a function of temperature.}
\label{fig:linearity}
\end{figure}
The gamma-ray photopeaks from $^{22}$Na (511 keV, 1275 keV) and $^{137}$Cs (662 keV) were fit with the triple-Gaussian plus linear background fit to determine the mean ADC channel of the primary photopeak component recorded by the CAEN digitizer.  The fits in ADC channel are plotted against their corresponding energies in Fig.~\ref{fig:linearity}.  The linearity over this energy range is excellent for all of the temperatures with $R^2$ values close to 1, showing that temperature has little effect on linearity in TLYC.  Based on previous results \cite{Hawrami2016}, excellent light-output linearity with energy is expected down to 60~keV.  The ADC channel is an uncalibrated measure of the integrated light output, therefore we can conclude that temperature does have a modest effect on the total light output.  As the temperature decreases the light output increases, with an observed change of $\sim$25\% from 50$^{\circ}$C to $-20^{\circ}$C, or 0.35\%/$^{\circ}$C.

\subsection{Energy Resolution}\label{sec:eres}

\begin{figure*}[!t]
\centering
\subfloat[\label{fig:fit_n20}]{\includegraphics[width=60mm]{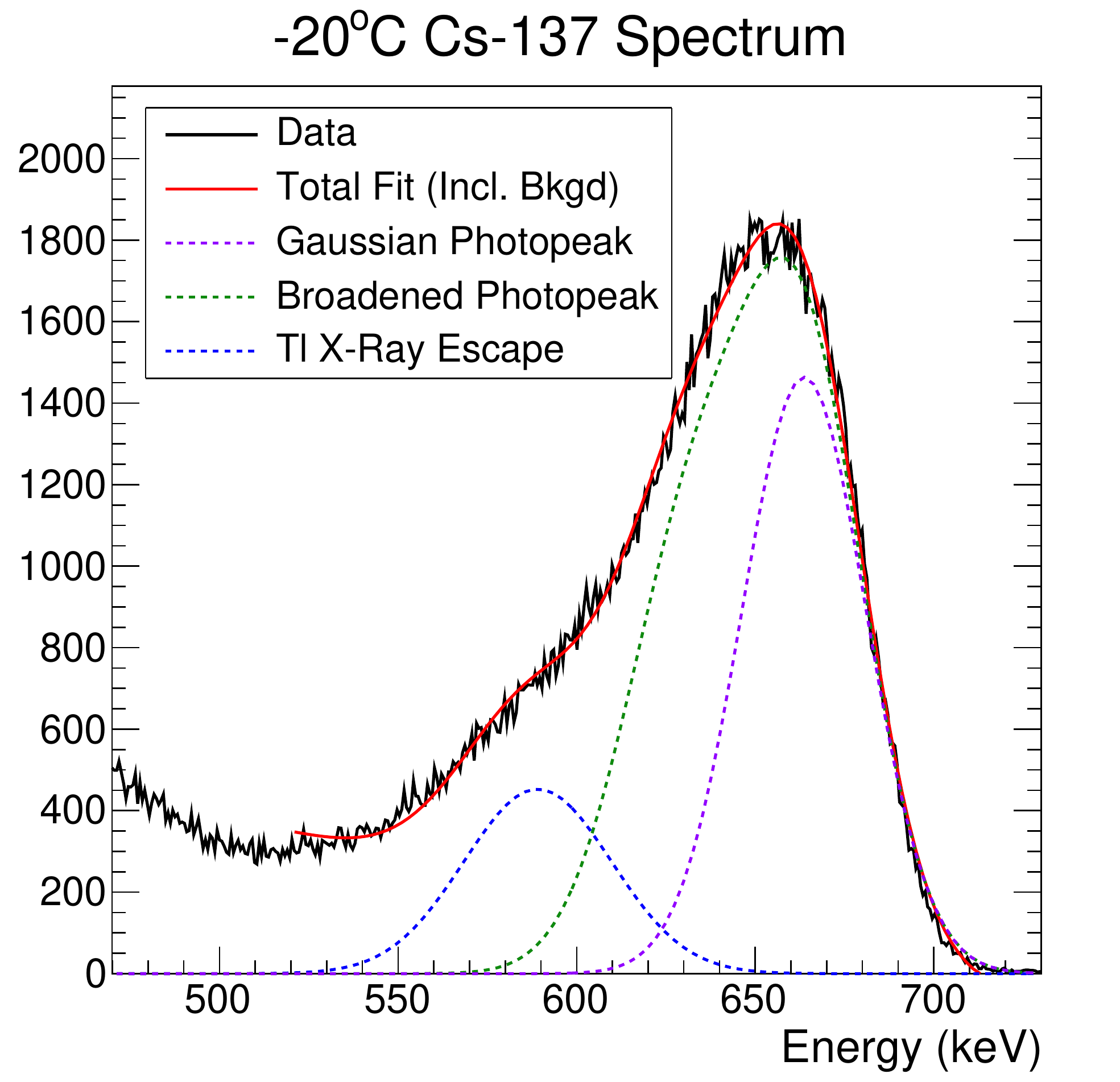}}
\subfloat[\label{fig:fit_10}]{\includegraphics[width=60mm]{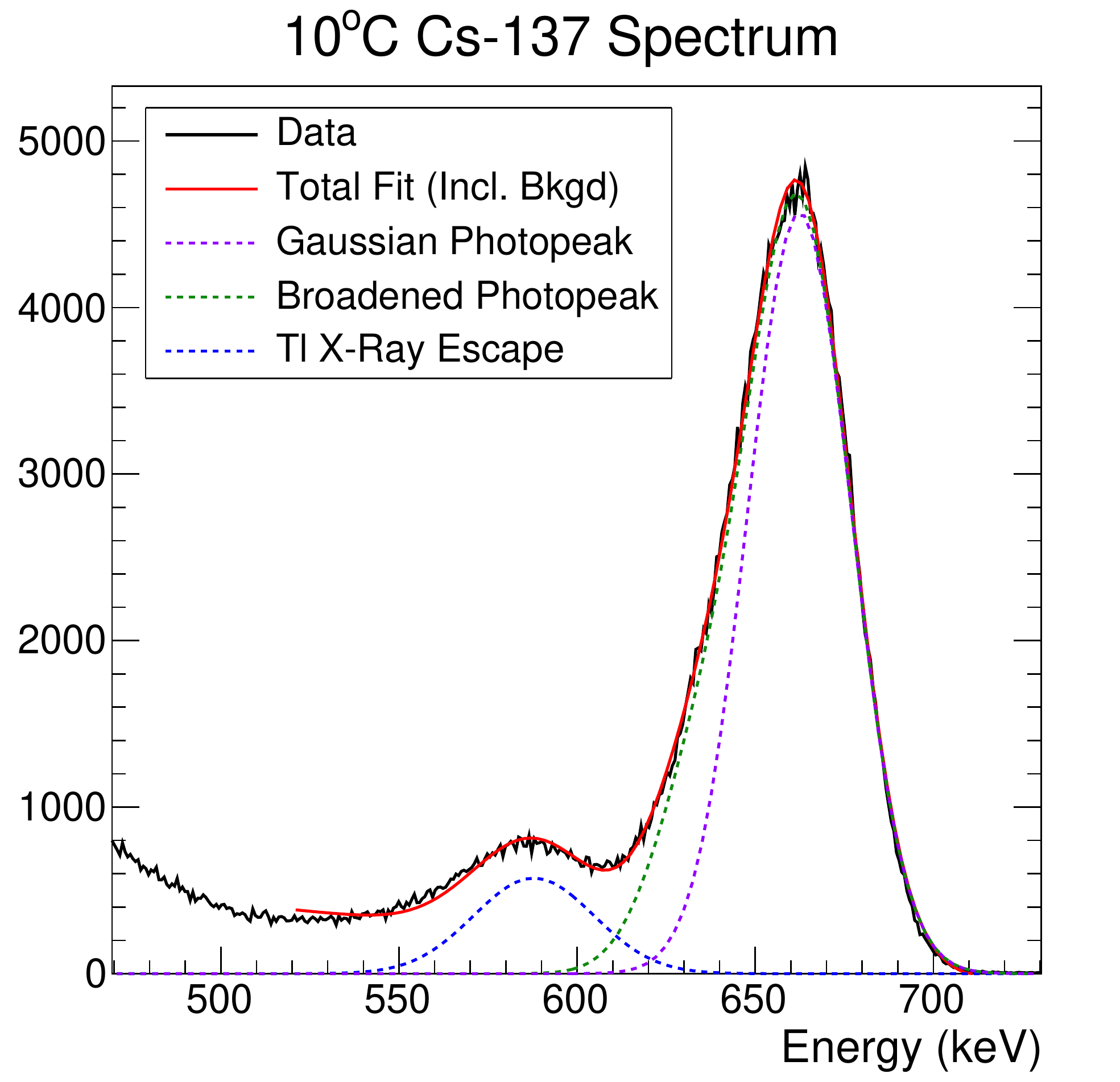}}
\subfloat[\label{fig:fit_50}]{\includegraphics[width=60mm]{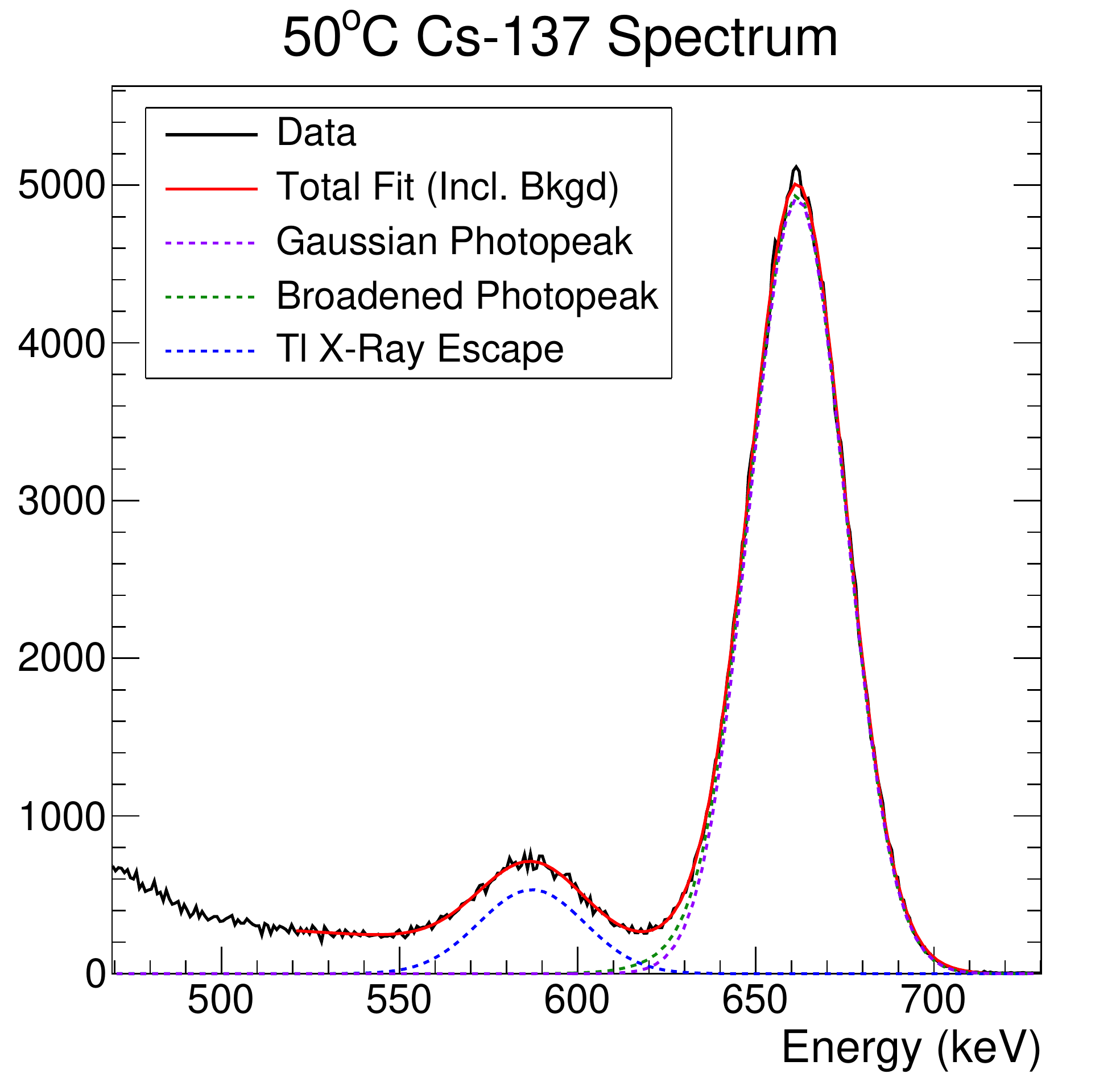}}
\caption{Triple-Gaussian plus linear background fits to the $^{137}$Cs spectrum at a) $-20^{\circ}$C, b) 10$^{\circ}$C, and c) 50$^{\circ}$C.}
\label{fig:fits}
\end{figure*}

\begin{figure}[b!]
\centering
\includegraphics[width=3.5in]{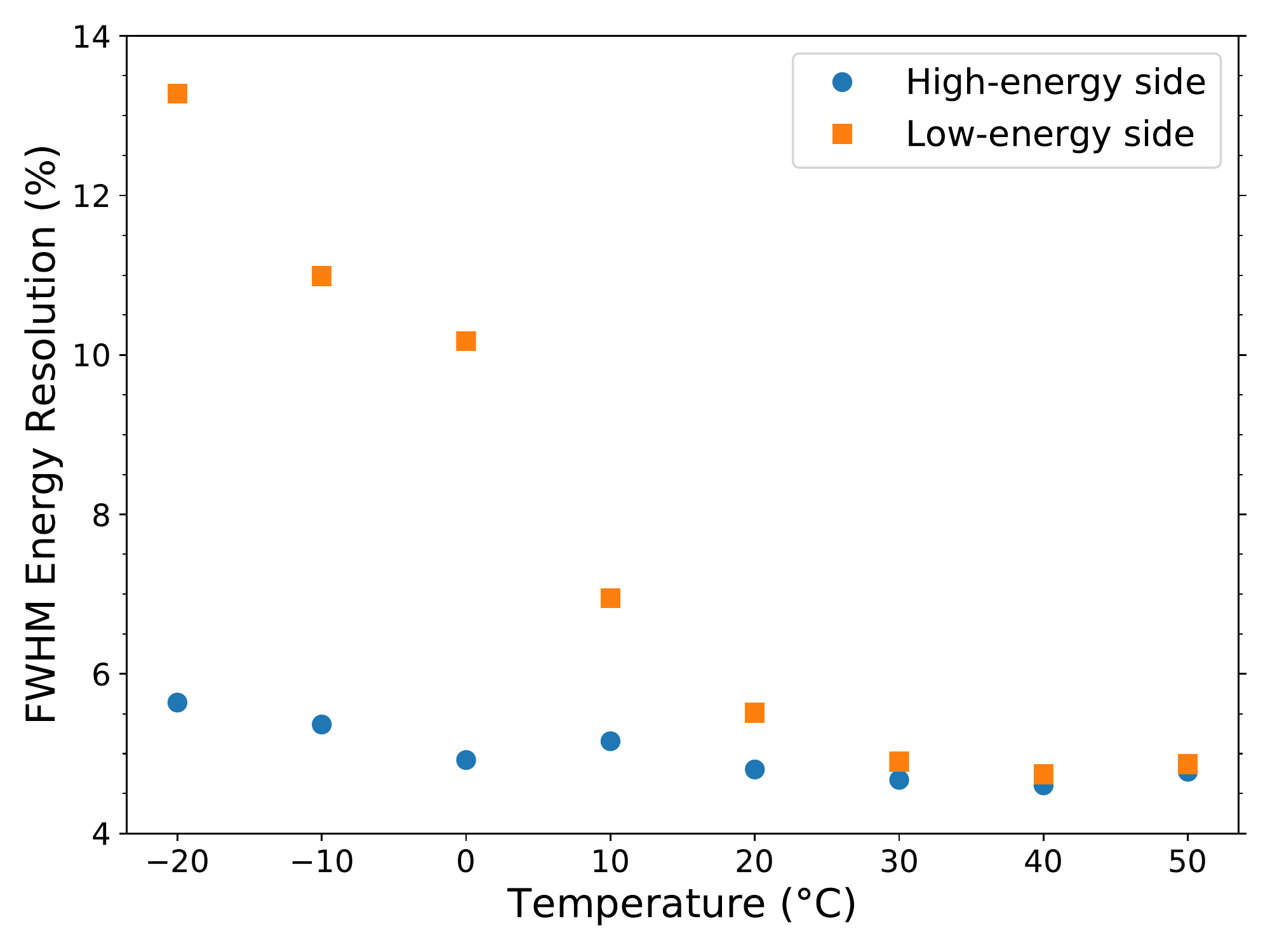}
\caption{The effective FWHM energy resolution at the 662~keV $^{137}$Cs photopeak as a function of temperature.}
\label{fig:energyres}
\end{figure}

The photopeak was observed to become asymmetrically broader and therefore the energy resolution degrades at lower temperature.  This effect can be seen in Fig.~\ref{fig:fits} which shows the triple-Gaussian plus linear background fit for the $^{137}$Cs spectrum at $-20^{\circ}$C, 10$^{\circ}$C, and 50$^{\circ}$C.  The FWHM energy resolution at the 662~keV $^{137}$Cs photopeak is shown as a function of temperature in Fig.~\ref{fig:energyres}.  Two sets of data are shown in Fig.~\ref{fig:energyres}.  Since the photopeak is not Gaussian, it cannot be described by a single FWHM in the standard way.  Instead, we calculate the half width half maximum from the mean of the broadened photopeak towards the high energy and low energy sides of the peak, and calculate an effective FWHM energy resolution based on each side as two times that value.  At high temperature these measures agree as the photopeak broadening is minimal, but as the temperature decrease these measures start to differ significantly.  The worsening energy resolution at lower temperatures is at odds with the increasing signal amplitude at those temperatures that was shown in Fig.~\ref{fig:linearity}. In modern scintillators, the most significant contribution to the energy resolution is often the event-to-event statistical variation in the electrons produced at the PMT photocathode. Larger pulse amplitudes indicate more photoelectrons and thus would be expected to produce a smaller statistical broadening of the peak. That the opposite result is observed indicates that there is a different contribution to the energy resolution that is temperature-varying.

Previously published results show that at room temperature TLYC's energy resolution is as good as 4.2\% for a similarly sized TLYC crystal to that used in this work \cite{Hawrami2016}.  The TLYC crystal sample used in this work with our experimental setup had an energy resolution of 5\% at room temperature.  Differences between our results and those previously published may be due to experimental setup, crystal quality, or Ce doping concentration, which is unknown for this sample.  At 50$^{\circ}$C the energy resolution is $\sim$4.8\%.  At $-20^{\circ}$C the measured energy resolution is 5.6\% based on the high-energy side of the photopeak and 13.3\% based on the low-energy side of the photopeak.

This broadening effect at low temperature was investigated by using several different measurement techniques: the spectra collected by the CAEN digitizer, integrating the waveforms collected by the Acqiris digitizer, and spectra collected using a shaped signal into a multichannel analyzer.  The broadening appeared in all measurements at $-20^{\circ}$C, therefore we determined it is not an artifact of the readout electronics.  We also verified that nearly all of the photopeak energy was captured by the integration windows used in the CAEN digitizer measurements.  Data with a longer integration window of 30~$\mu$s were acquired at $-20^{\circ}$C in a separate run after the initial thermal cycling study and the asymmetric broadening was still present, therefore the effect is not likely due to partial light collection.  The crystal and PMT were brought to room temperature, decoupled, and recoupled before the measurement with the longer integration window, therefore the effect is not from an issue with the optical coupling.  We also verified that the crystal performance at room temperature had returned to the pre-thermal cycling performance after being at -20$^{\circ}$C.  In addition, an asymmetric broadening of the neutron peak as temperature decreased was also observed.  The observation of this effect in both the gamma-ray and neutron signals suggests a common origin and may point to properties of the crystal or light output production and propagation as possible causes.

\subsection{Neutron Energy}      

\begin{figure}[h]
\centering
\includegraphics[width=3.5in]{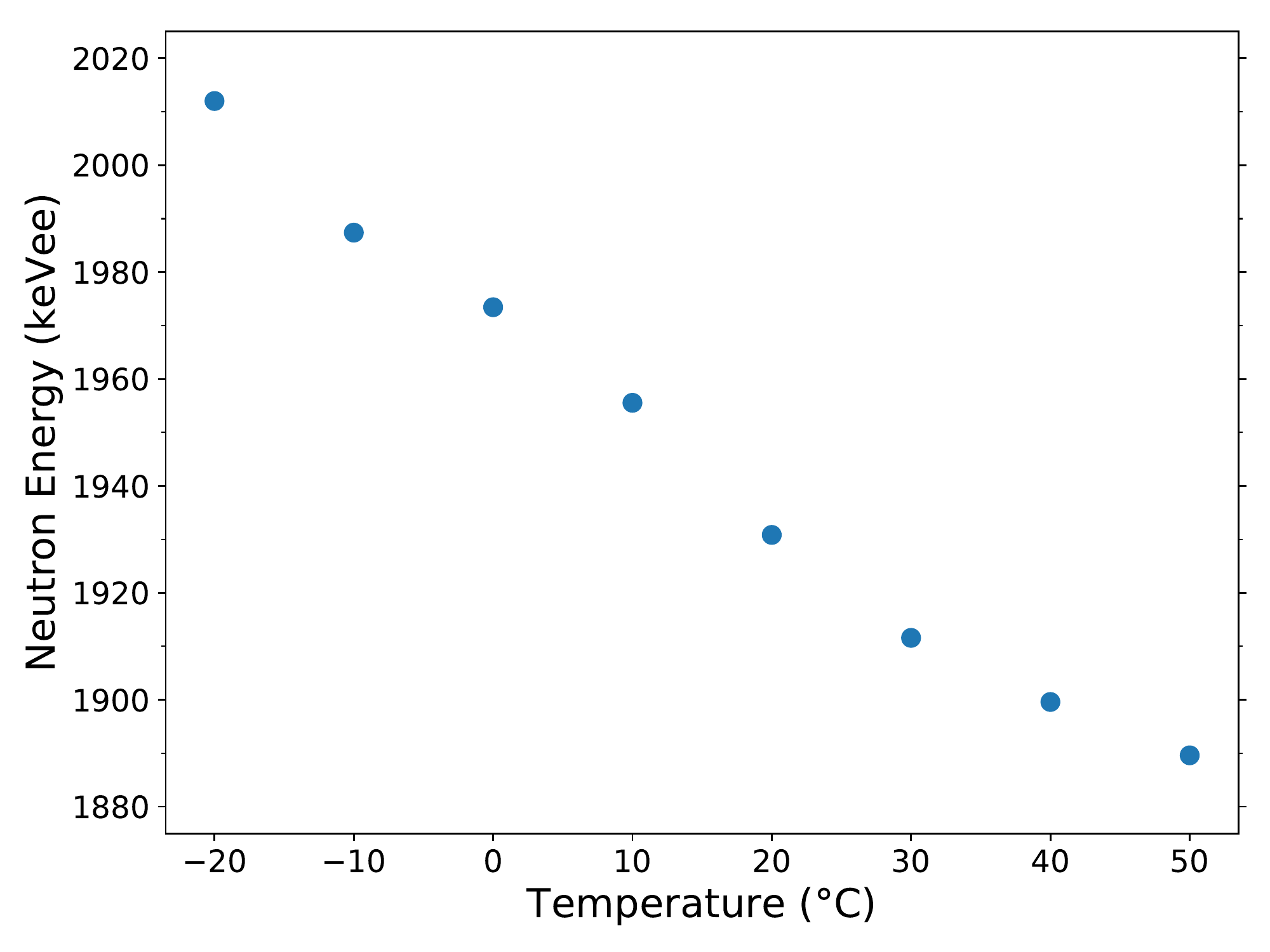}
\caption{Calculated energy of the neutron capture as a function of temperature.}
\label{fig:neutrons}
\end{figure}

The detected energy of the neutron capture reaction was calculated by fitting the neutron peak in the $^{252}$Cf spectrum with a Gaussian plus linear background.  The mean ADC channel was then calibrated into gamma-equivalent energy with a linear conversion based on the gamma-ray peaks determined in Section~\ref{sec:lin}.  Because of inefficiency in the light output generated by the heavy particles of the neutron capture reaction, the 4.8 MeV Q value of the reaction is detected at a quenched energy.  Fig.~\ref{fig:neutrons} shows the detected neutron energy versus temperature.  There is a negative correlation between temperature and the detected neutron energy.  However, the change in the energy of the detected neutrons is small. There is only a 6.5\% increase in the detected energy of the neutrons at $-20^{\circ}$C (2012 keVee) relative to 50$^{\circ}$C (1890 keVee), showing that temperature does not have a significant impact on the neutron energy.  These measurements are in agreement with previous measurements that reported $\sim$1900 keVee at room temperature \cite{Hawrami2016}.

\subsection{Figure of Merit}\label{sec:fom}

\begin{figure}[t]
\centering
\subfloat[\label{fig:optfom_a}]{\includegraphics[width=3.5in]{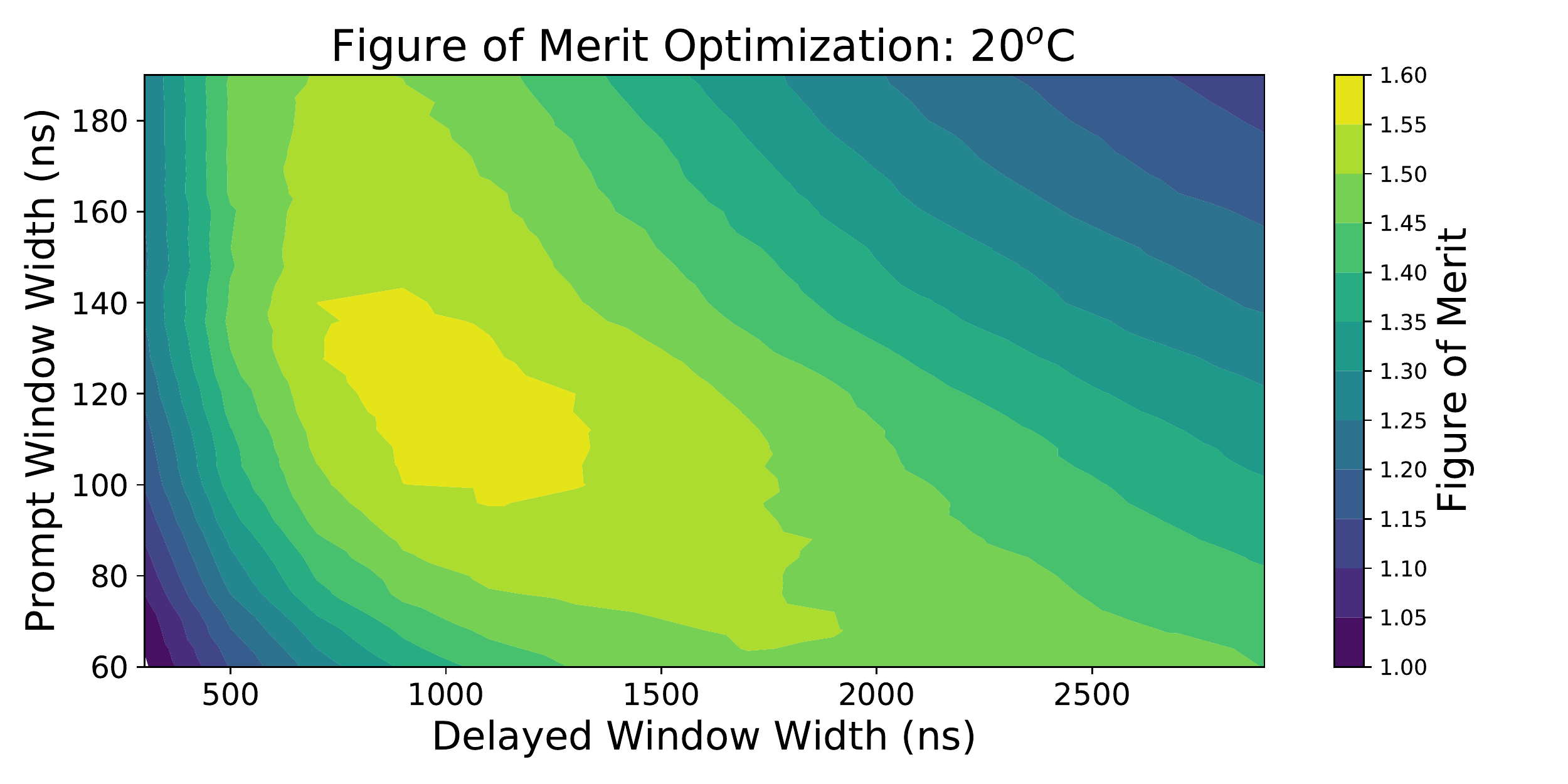}}\\
\subfloat[\label{fig:optfom_b}]{\includegraphics[width=3.5in]{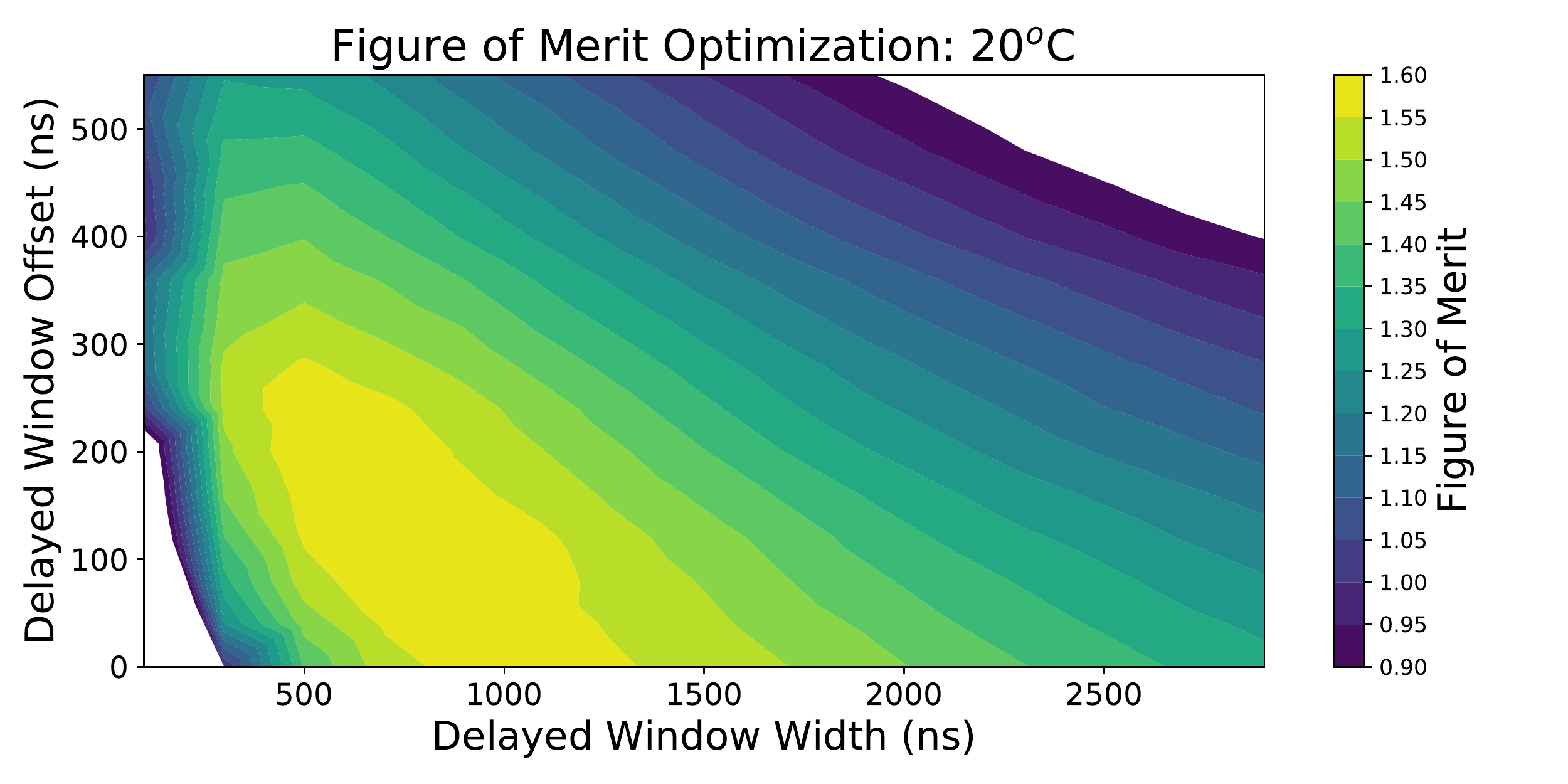}}
\caption{Example FOM results from the window optimization routine described in the text, for 20$^{\circ}$C.}
\label{fig:optfom}
\end{figure}

\begin{figure}[h!]
\centering
\includegraphics[width=3.5in]{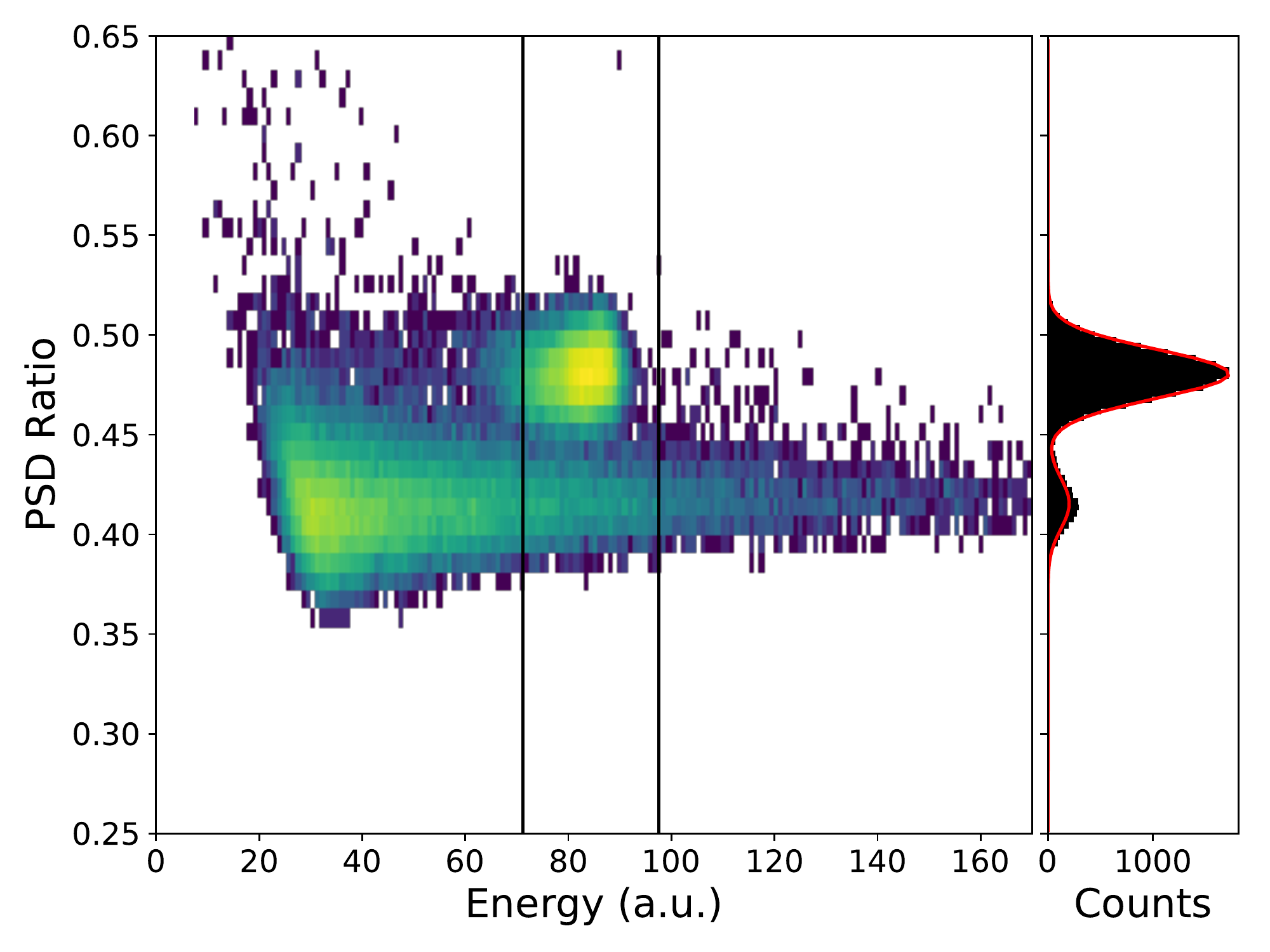}
\includegraphics[width=3.5in]{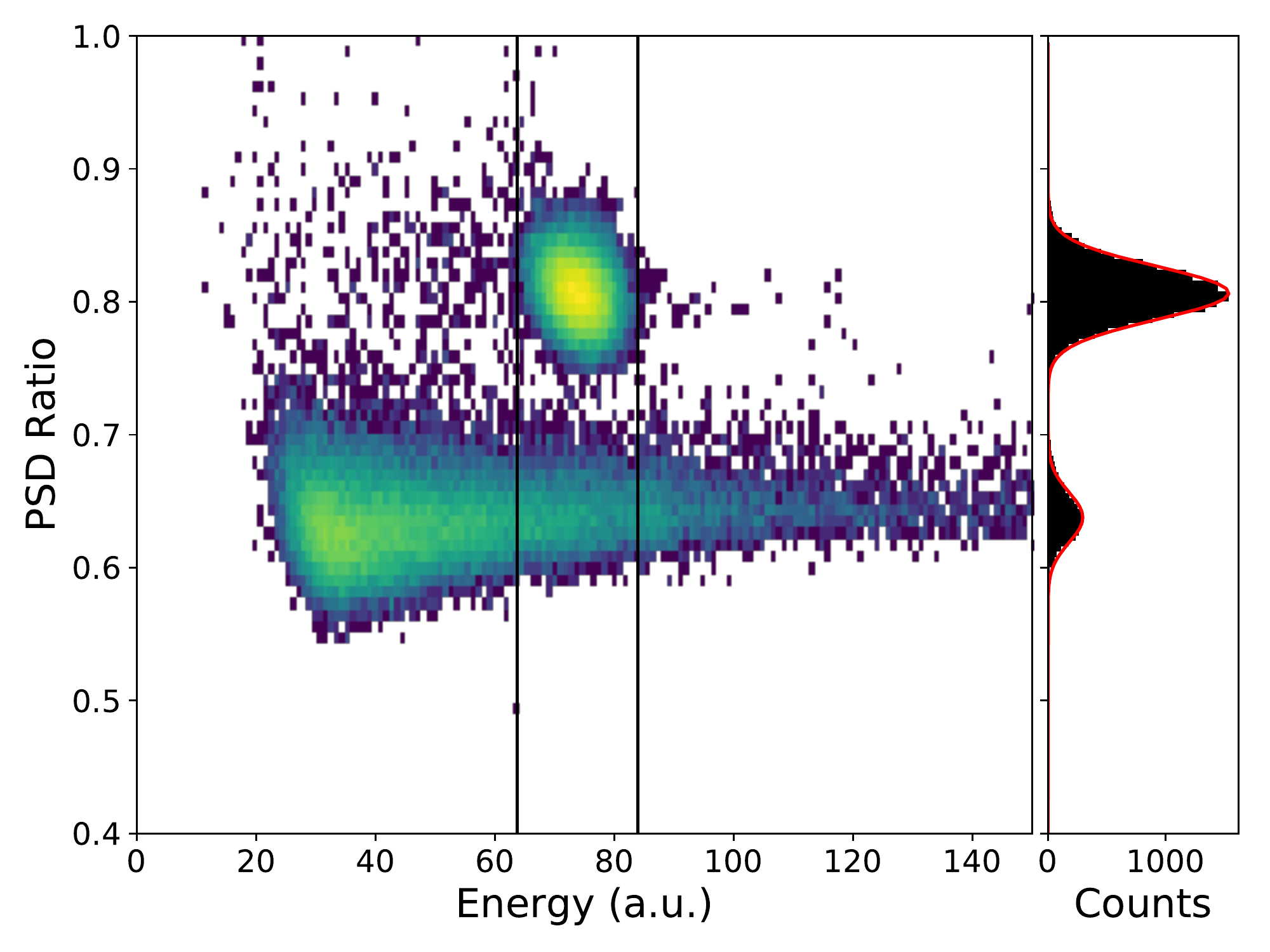}
\caption{PSD Ratio versus energy using optimized windows at $-20^{\circ}$C (top) and 50$^{\circ}$C (bottom).  The neutrons appear as the blob with higher a PSD ratio and are separated from the continuous gamma-ray line at lower PSD ratio.  The right panels show the corresponding histogram of PSD ratio within $\pm$3$\sigma$ of the neutron peak (vertical black lines) with fit used to calculate FOM (red).}
\label{fig:fomplot}
\end{figure}

The prompt and delayed integration windows were used to calculate the PSD ratio ($P/D$) and the FOM (\ref{eq1}).  A cut in energy of $\pm3\sigma$ around the neutron peak based on a Gaussian fit was applied.  Two sets of windows were used to compare the FOM as a function of temperature.  The first set of windows were determined at room temperature to maximize the FOM requiring the prompt and delayed windows were adjoined and remained fixed for all temperatures.   The second set of windows were determined for each temperature by an optimization routine meant to provide a consistent method of determining windows across temperatures.  The first step of the optimization routine was to calculate the FOM for a range of prompt and delayed windows, assuming the windows were adjoined (\textit{e.g.} Fig.~\ref{fig:optfom_a}).  The prompt window where the best FOM was obtained from this step was fixed in the subsequent step where the width of the delayed window and an offset between the prompt and delayed windows were varied (\textit{e.g.} Fig.~\ref{fig:optfom_b}).  There is only a small increase in FOM when an offset between the prompt and delayed windows is used.  Examples of the optimized FOM at $-20^{\circ}$C and 50$^{\circ}$C are shown in Fig.~\ref{fig:fomplot}, along with the PSD ratio used to calculate the FOM.

\begin{figure}[h!]
\centering
\includegraphics[width=3.5in]{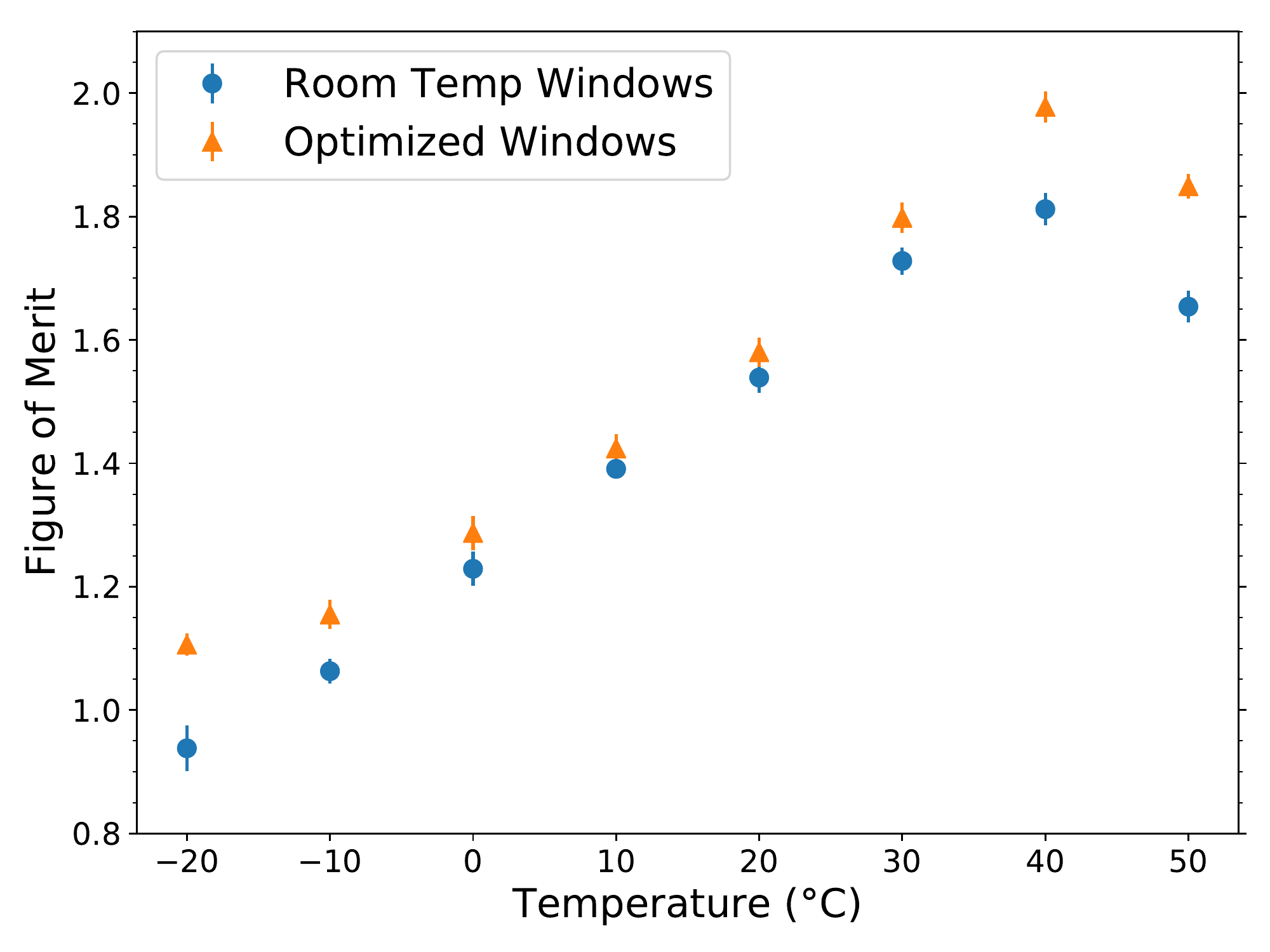}
\caption{Figure of Merit as a function of temperature for two sets of windows.}
\label{fig:fom}
\end{figure}
Previously published results on TLYC show a FOM as good as 2 at room temperature for a similarly sized TLYC crystal to that used in this work \cite{Hawrami2016}, slightly better than the FOM of 1.6 measured in this work at 20$^{\circ}$C.  The PSD ratio was defined differently in \cite{Hawrami2016}, as $D/(P+D)$, however, the same FOM is calculated from each method.  The FOM as a function of temperature is shown in Fig.~\ref{fig:fom}.  As the temperature increases the FOM generally increases.  There is a slight improvement at the temperature extremes if optimized windows are used at each temperature, however, even with fixed windows the FOM is always above 0.9.  The highest FOM ($\sim$2) was measured at 40$^{\circ}$C.  At 50$^{\circ}$C the neutron peak becomes broader in PSD ratio, causing the FOM to be slightly worse.

\subsection{Pulse Shapes}

\begin{figure}[h]
\centering
\includegraphics[width=3.5in]{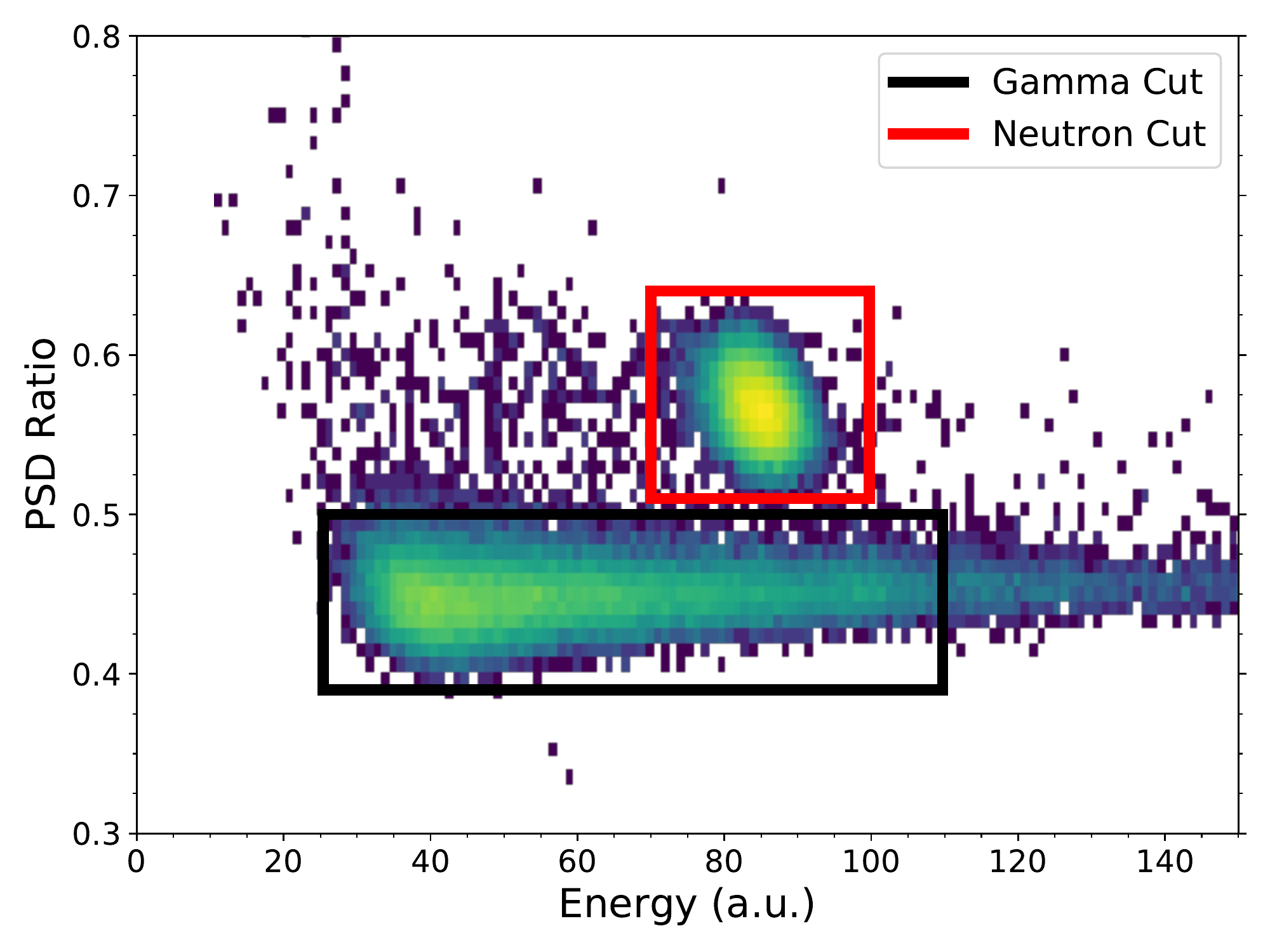}
\caption{Example FOM plot at 20$^{\circ}$C with energy and PSD cut regions used to determine average waveforms indicated.}
\label{fig:fomcut}
\end{figure}

\begin{figure}[b!]
\centering
\includegraphics[width=3.5in]{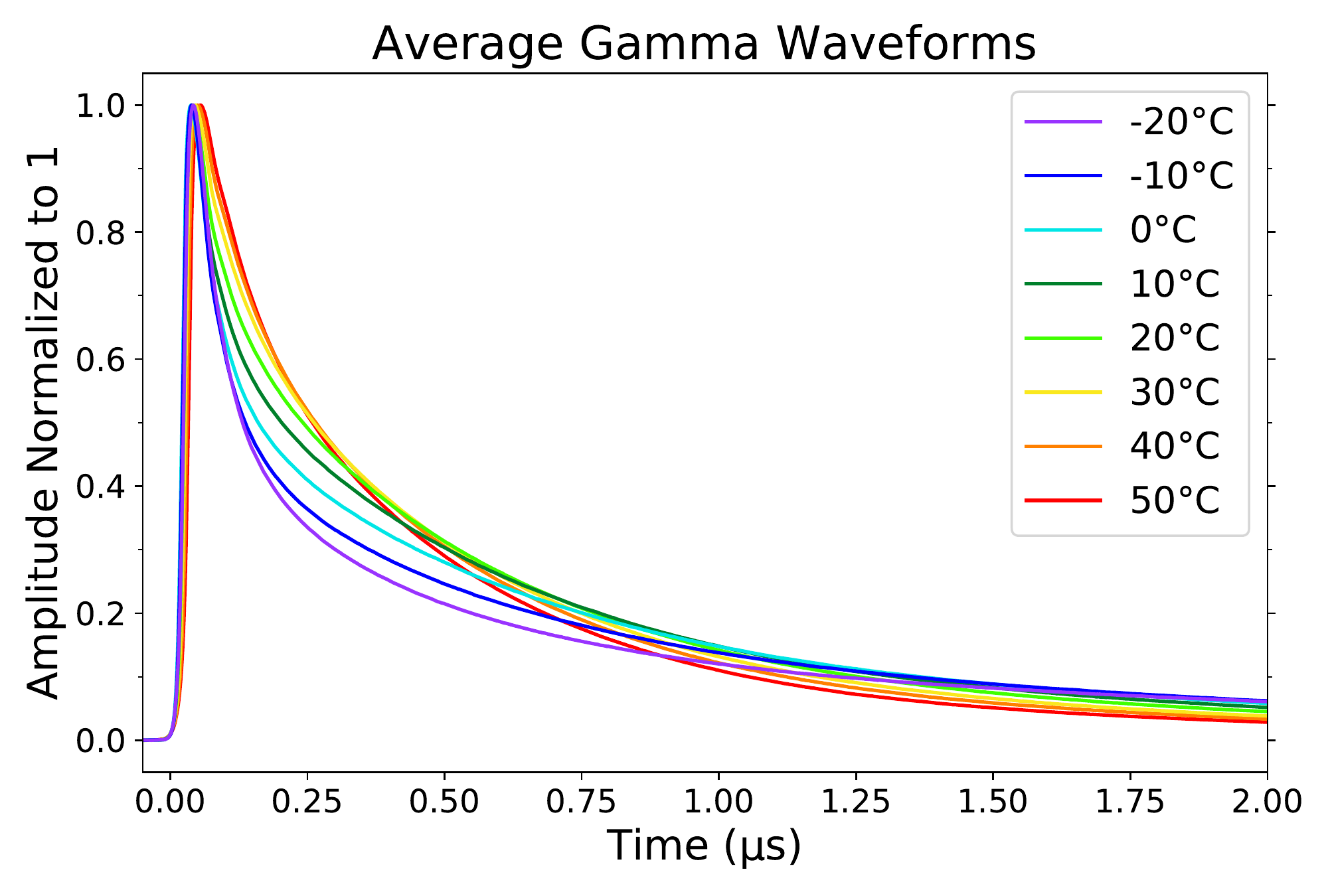}
\includegraphics[width=3.5in]{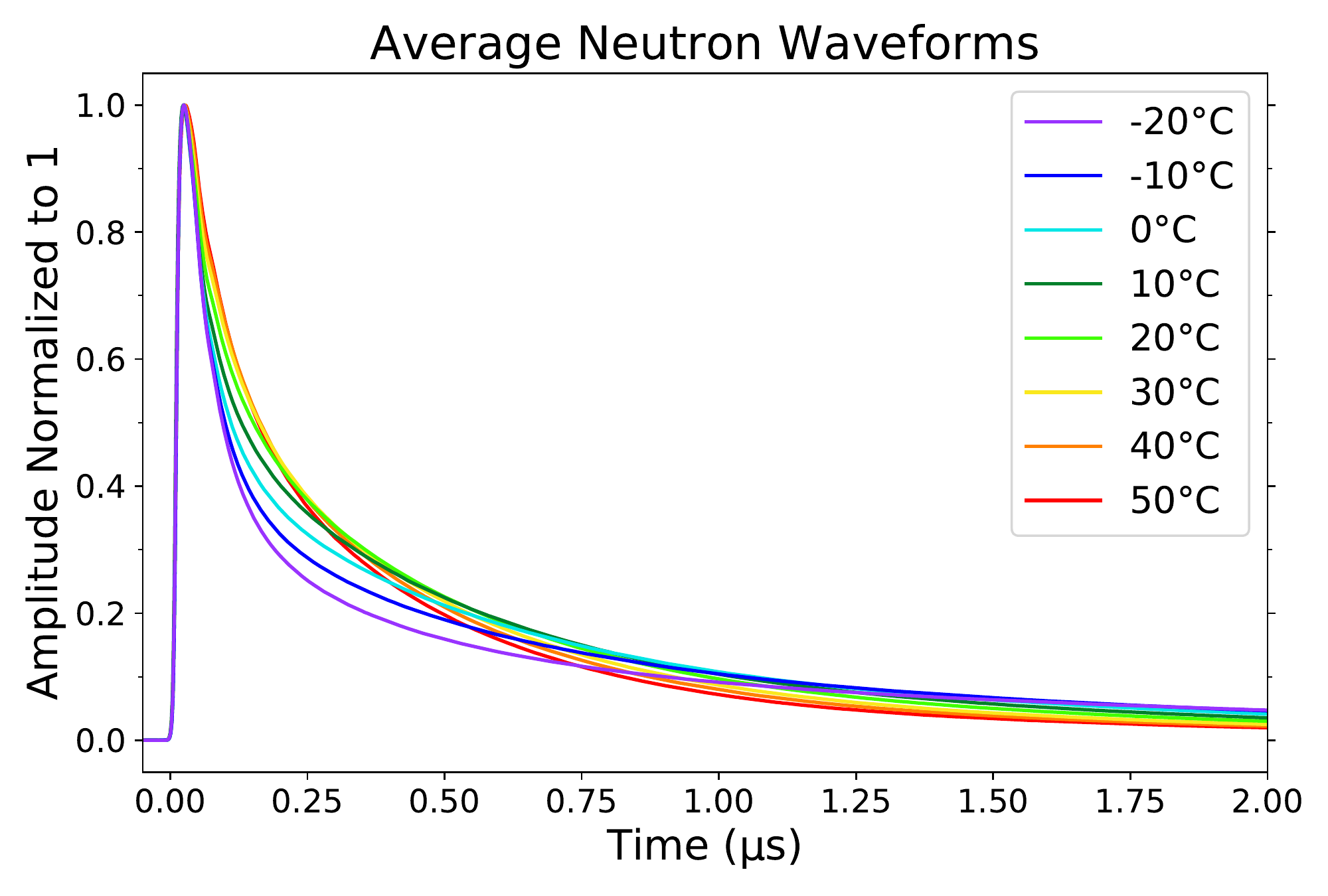}
\caption{Gamma (top) and neutron (bottom) waveforms at the range of temperatures measured, normalized to unity.}
\label{fig:wavenorm1}
\end{figure}

The optimized PSD ratio from Sec.~\ref{sec:fom} was used and energy and PSD cuts made to isolate the gammas from the neutrons as illustrated in Fig.~\ref{fig:fomcut}.  From these cuts the average waveforms for neutrons and gammas were extracted.  Fig.~\ref{fig:wavenorm1} shows the average waveforms normalized to unity to see the differences in the pulse time profile.  As the temperature increases the pulse decay times at early times become longer and at long times become shorter.  

\begin{figure}[t]
\centering
\includegraphics[width=3.5in]{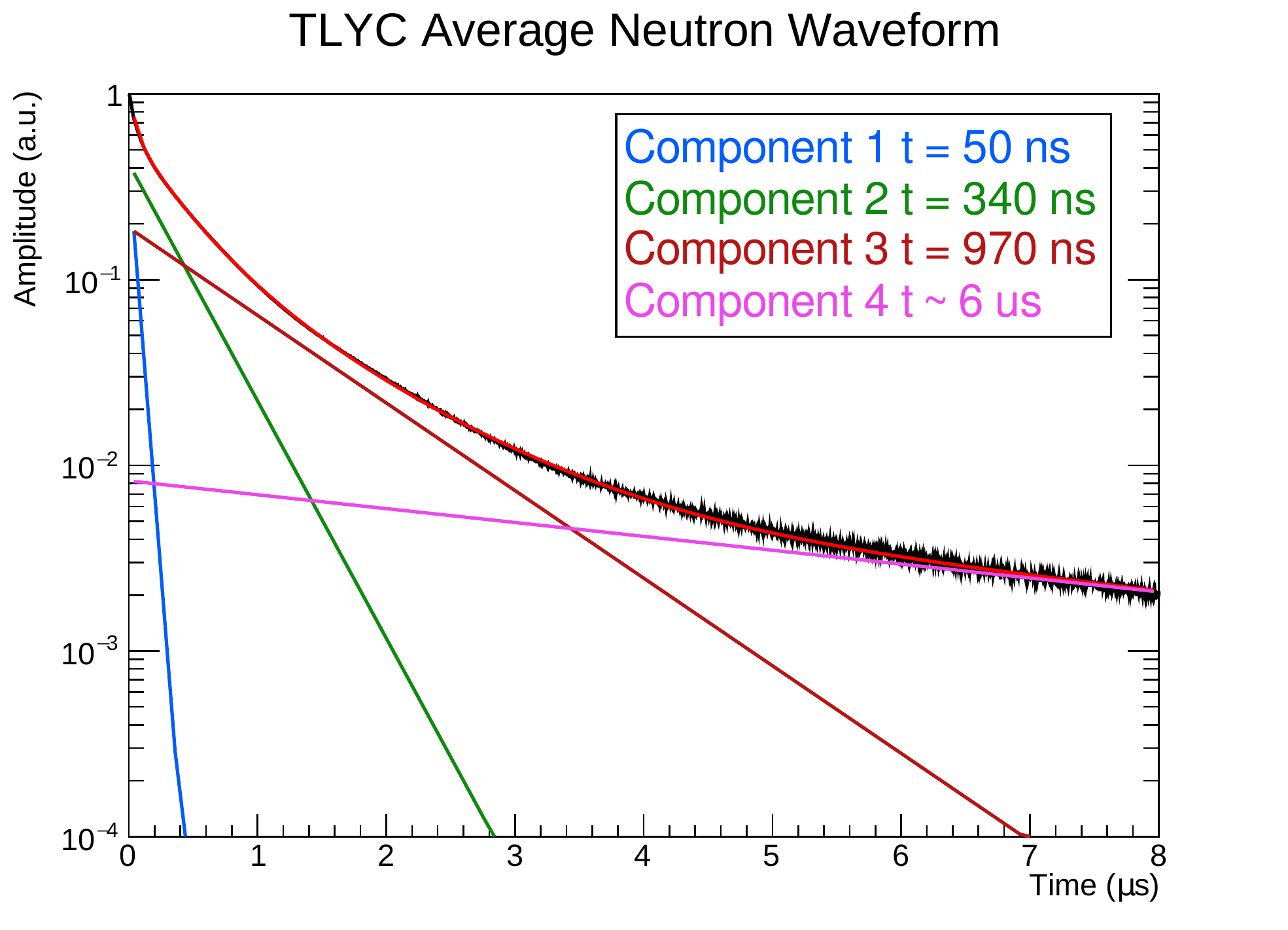}
\caption{Example of decay times fitted to the average neutron waveform at 20$^{\circ}$C.}
\label{fig:decayex}
\end{figure}
\begin{figure}[b!]
\centering
\includegraphics[width=3.5in]{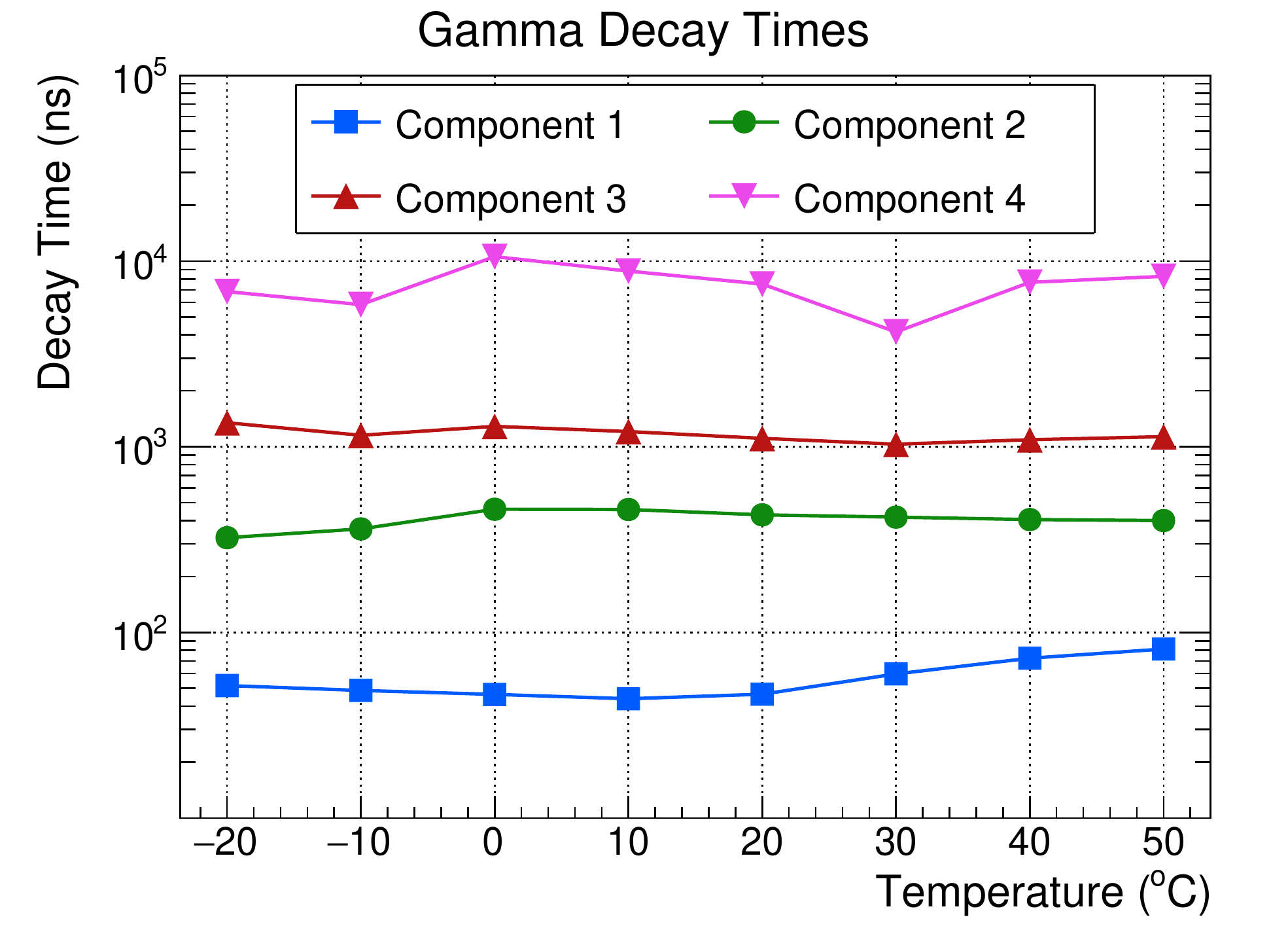}
\includegraphics[width=3.5in]{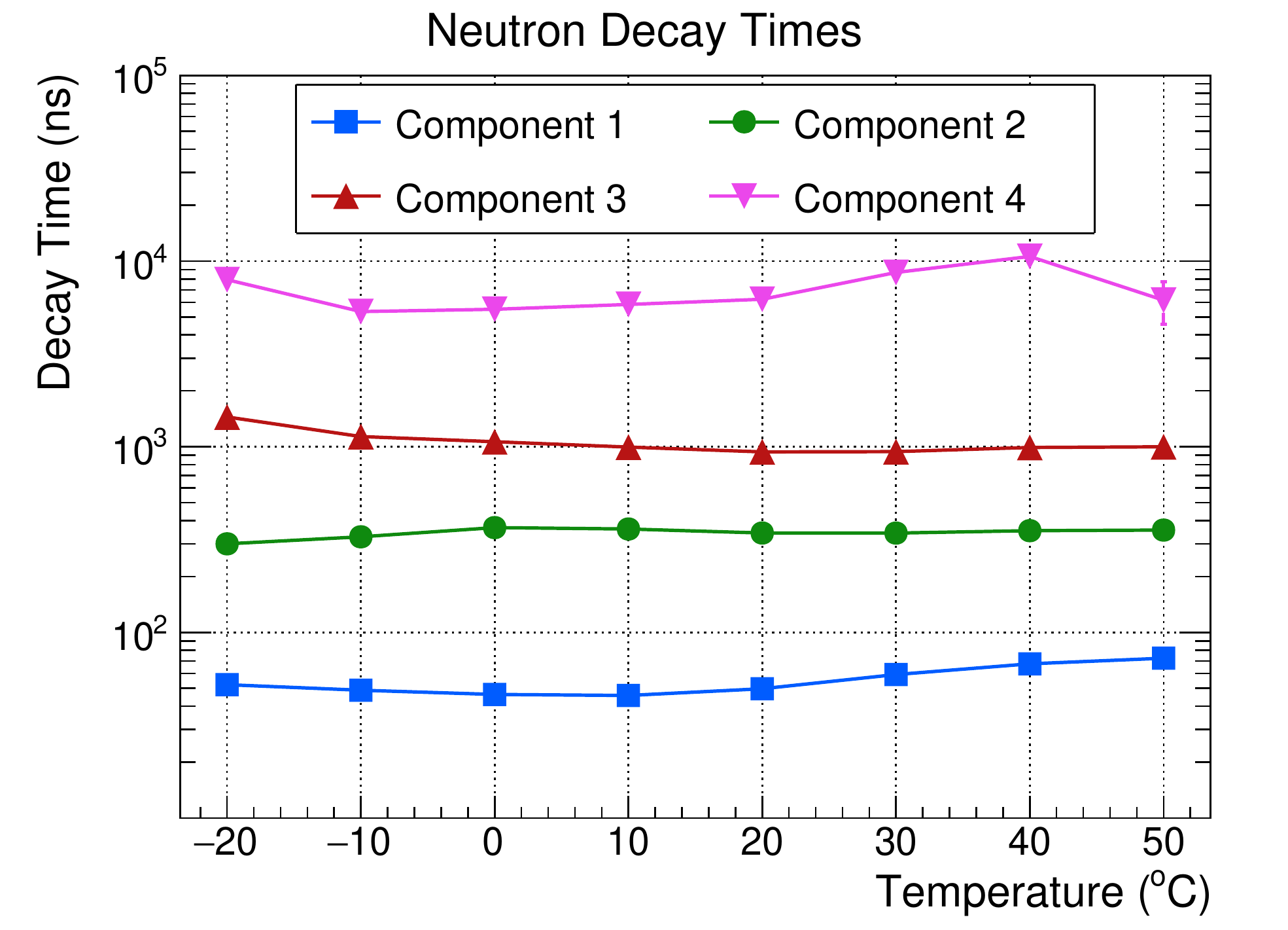}
\caption{Fitted decay times of the four decay components of the gamma (top) and neutron (bottom) waveforms as a function of temperature.}
\label{fig:decaytimes}
\end{figure}

To further illustrate this, the average pulse shapes were fit with a sum of exponential functions $(\sum_{n} A_ne^{-t/\tau_n})$ to extract decay time components.  An example of the 4-component fit is shown in Fig.~\ref{fig:decayex}.  Note that previous publications \cite{Kim2016,Hawrami2016,Hawrami2017} report 3 decay times; these are consistent with the three early components of the fits used here.  The longest component is $\sim$6--8~$\mu$s and remains fairly stable over temperature.  The decay times measured near room temperature are in good agreement with previous publications, with a short component of $\sim$50~ns that is similar for the neutron and gamma waveforms, and longer components of $\sim$350~ns and $\sim$970~ns for the neutron waveforms and $\sim$430~ns and $\sim$1100~ns for the gamma waveforms.  With these decay times and as shown in Fig.~\ref{fig:light}, the gamma pulses are slower than the neutron pulses.

\begin{figure}[b!]
\centering
\includegraphics[width=3.5in]{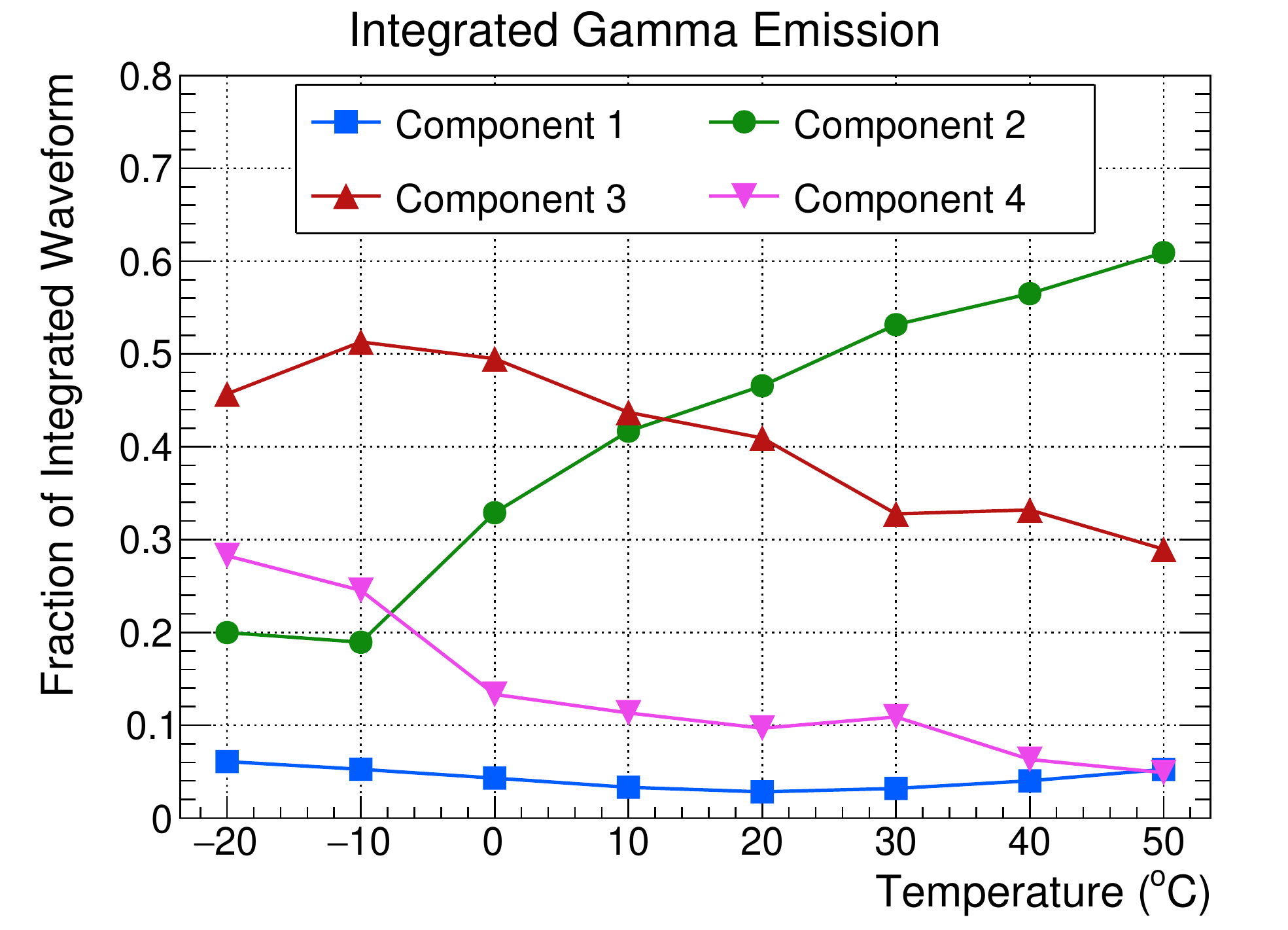}
\includegraphics[width=3.5in]{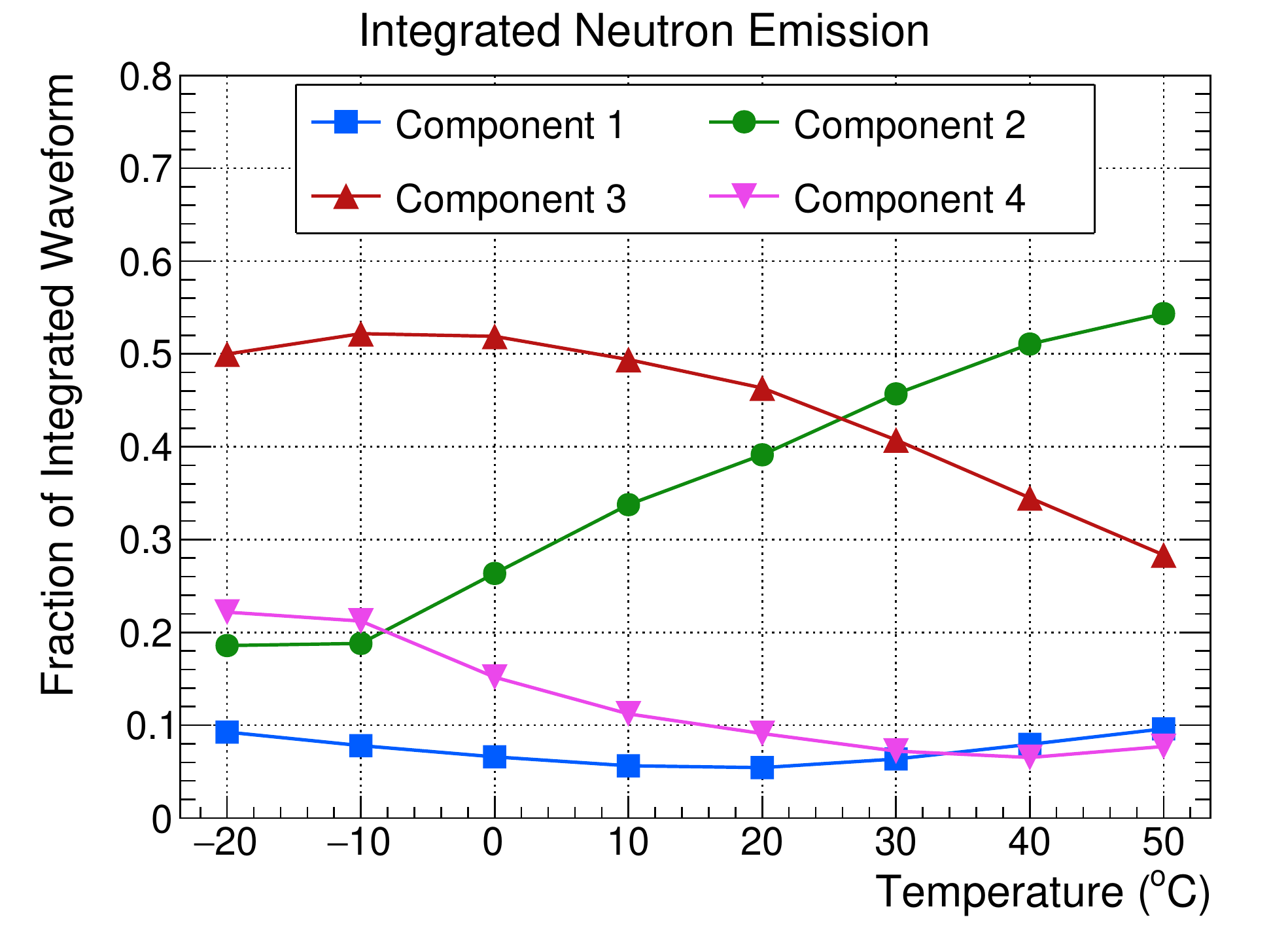}
\caption{Integrated emission of the four decay components to the gamma (top) and neutron (bottom) waveforms as a function of temperature.}
\label{fig:decaytimeint}
\end{figure}
Fig.~\ref{fig:decaytimes} shows the four decay times as a function of temperature.  Fitting decay times is difficult, as the uncertainties can be large and the parameters highly correlated, so this plot is intended to be used to show the relative trends in how the decay times change with temperature, rather than providing absolute numbers at each temperature.  A more meaningful quantity to compare is the integrated contribution of each component to the total integrated waveform (from 0--8~$\mu$s), which is shown in Fig.~\ref{fig:decaytimeint}.
The two fastest components have fairly stable decay times with temperature.  However, the integrated contribution of the second component decreases with decreasing temperature, from about $\sim$60\% at $50^{\circ}$C to $\sim$20\% at $-20^{\circ}$C, while the integrated contribution of the fastest component does not strongly correlate with temperature.  The fastest component also contributes least to the total integrated waveform, $<$10\%.  The decay time of component 3 increases some with decreasing temperature, more so for the neutron waveforms.  Correspondingly, the contribution of this component to the integrated waveform increases as the temperature decreases, from $\sim$30\% at 50$^{\circ}$C to $\sim$50\% at $-20^{\circ}$C.  The contribution of the longest decay component also increases with decreasing temperature, but with a lower overall integrated contribution relative to component 3.  

\begin{figure}[h]
\centering
\includegraphics[width=3.5in]{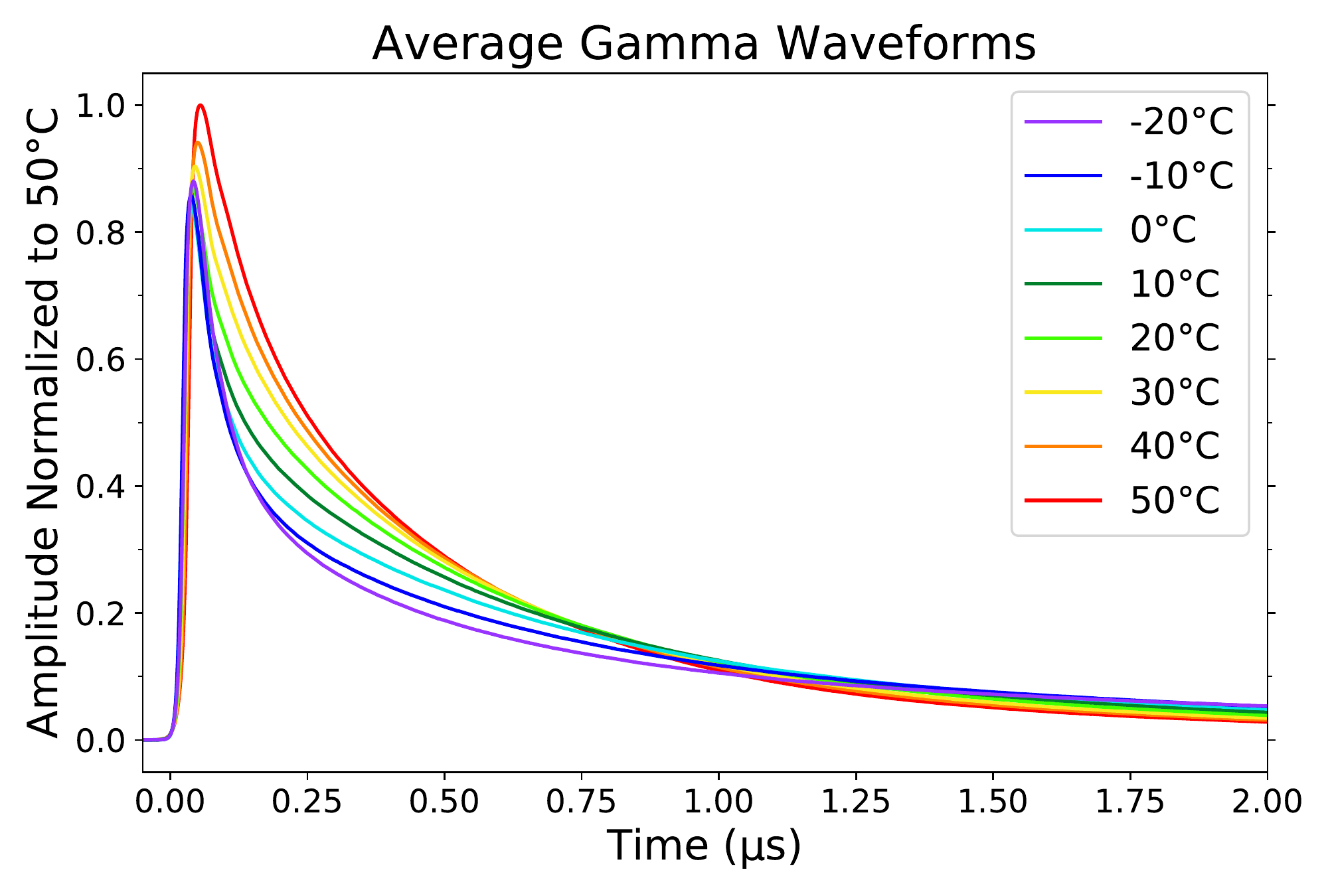}
\includegraphics[width=3.5in]{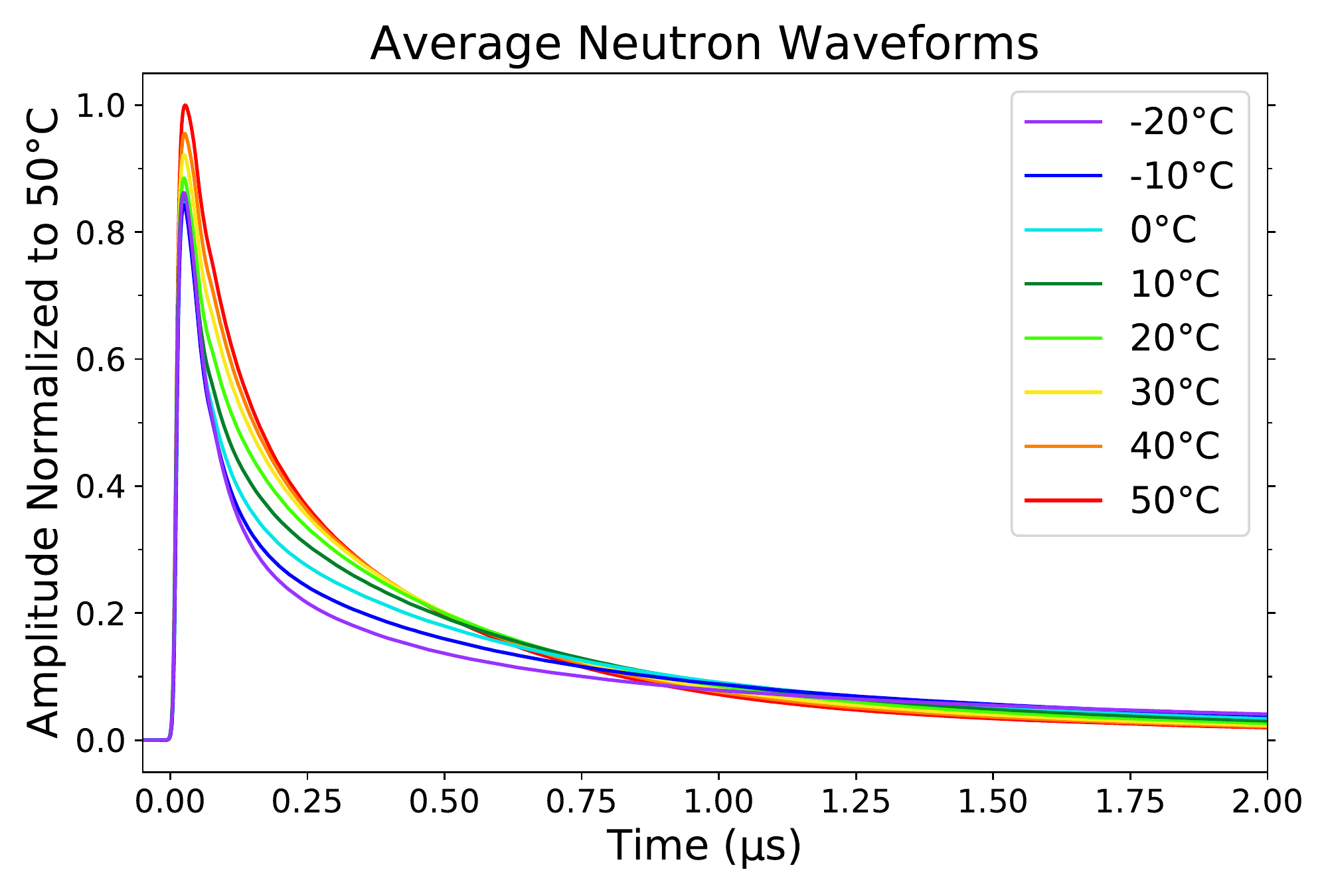}
\caption{Gamma (top) and neutron (bottom) waveforms at the range of temperatures measured, normalized to the 50$^{\circ}$C peak amplitude.}
\label{fig:wavenorm50}
\end{figure}

Fig.~\ref{fig:wavenorm50} shows the average waveforms normalized to 50$^{\circ}$C, which had the largest peak amplitude.  Contrary to the total light output increasing with decreasing temperature as presented in Sec.~\ref{sec:lin} and shown in Fig.~\ref{fig:linearity}, the pulse amplitude actually decreases with temperature by $\sim$15\% from 50$^{\circ}$C to $-20^{\circ}$C.  Combined with the decrease in decay times at early times as the temperature decreases, this confirms that an increasing fraction of the light output is coming from long times in the tail of the pulse as the temperature decreases.

\subsection{Comparison to Selected Elpasolites}

The TLYC results indicate both similarities and differences to published thermal characterizations of CLYC \cite{Budden2012,Menge2011} and CLLBC \cite{Coupland2017}. In all three of these elpasolite scintillators, the calibrated energy of the neutron peak decreases with increasing temperature, a result of thermally-dependent scintillation efficiency of the alpha and triton produced by the $^{6}$Li neutron capture. As previously reported at room temperature, the relative light output of the neutron peak is much lower in TLYC than in the other common elpasolites, with an electron-equivalent energy of 1.9 MeV for TLYC compared to 3.2 for CLYC and 3.0 for CLLBC.

The fits to the TLYC pulse shapes can be compared to previous analysis of CLYC pulse shapes to inform the interpretation of the multiple decay components. The short TLYC component (component 1, $\sim$50 ns) is similar in magnitude to the 72 ns component of CLYC measured and attributed to direct Ce$^{3+}$ emission, but is only observed for gammas in CLYC \cite{Budden2012}. Component 2 (350 ns for gammas and 430 ns for neutrons) is similar in magnitude to the intermediate decay time component in CLYC associated with V$_k$ formation and migration, and the longer components (3 and 4) are similar in magnitude to the slow Ce$^{3+}$ capture due to the formation of self-trapped exitons (STE) identified in CLYC (few $\mu$s decay time) \cite{Budden2012}. Both TLYC and CLLBC lack the ultra-fast core-to-valence luminescence (CVL) component that is present in CLYC gamma waveforms; CLYC is also the only scintillator of the three where the gamma pulses are faster than neutron pulses and where the neutron and gamma waveforms differ most in the prompt PSD window rather than the delayed window.

In all three elpasolite scintillators mentioned, the light pulse returns to baseline faster at higher temperatures. In CLYC, this effect was identified with thermally-activated motion of the V$_k$ and STE charge carriers to the Ce-doping centers. In CLYC, the relative contribution of the CVL to the gamma waveforms decreases with increasing temperature, correlated with a decreasing PSD FOM from 4.2 at $-20^{\circ}$C to 1.8 at 50$^{\circ}$C \cite{Budden2012}. The opposite PSD trend is observed in CLLBC \cite{Coupland2017} and the current TLYC results, in which the FOM increases with increasing temperature. This is likely associated with the fundamental differences in waveform components and relative differences between neutron and gamma waveforms mentioned in the previous paragraph.

\section{Summary \& Conclusion}

TLYC is a relatively new elpasolite scintillator under study that can provide superior gamma-ray photopeak detection efficiency over the more-mature CLYC.  TLYC's performance was tested for the first time over a temperature range from $-20^{\circ}$C to 50$^{\circ}$C in 10$^{\circ}$C steps.  At all of the temperatures TLYC maintains linearity over the energy range tested with little change in slope between temperatures.  The light output as measured by gamma-ray photopeaks increases as the temperature decreases.  However, the best resolution of 4.6\% was measured at 40$^{\circ}$C.  The energy resolution of TLYC is worse at low temperatures due to an observed asymmetric broadening of the photopeak to low energy, and is 5.6\% calculated using the high-energy side of the photopeak and 13.3\% using the low-energy side of the photopeak, that is most effected by the broadening, at $-20^{\circ}$C. The energy of the neutron capture peak is minimally affected by temperature.  Contrary to the light output, the pulse heights increase with temperature, but changes in component decay times and amplitudes with temperature lead to higher light output as temperature decreases with an increasing contribution to the integrated waveform from components with longer decay times.  The figure of merit increases with temperature from 1.1 at $-20^{\circ}$C to 2.0 at 40$^{\circ}$C.  At 50$^{\circ}$C the performance is slightly worse than 40$^{\circ}$C, possibly due to broadening of the neutron peak.

Our study used only a single relatively small TLYC sample. Performance is not expected to vary considerably with crystal size, since self-absorption is not a large effect; this can be seen in room-temperature characterizations of a larger TLYC crystal \cite{Hawrami2017} and expectations from weak dependence of CLYC crystal performance on size up through a 3'' right cylinder \cite{SoundaraPandian17}. The detection efficiency will of course be larger for larger detectors, and the contribution of the X-ray escape peak observed in TLYC will be smaller. Comparing TLYC to CLYC and CLLBC shows that all of these elpasolite scintillators have strengths and weaknesses relating to detection that can influence the choice of material for a particular application. For example, TLYC and CLLBC have an opposite PSD performance trend with temperature compared to CLYC, so that if PSD at low temperature is important then CLYC is a better option. CLLBC has better gamma-ray energy resolution than CLYC or TLYC but often has alpha contamination that can interfere with low-rate neutron detection. TLYC has significantly better gamma-ray photopeak detection efficiency than the other two materials but the prominent X-ray escape peak complicates spectral unfolding.

Our study also did not examine the origin of the observed temperature-dependent asymmetric broadening of the photopeak, except to rule out artifacts due to readout electronics, integration time, optical coupling, and thermal damage.  The observed effect could result from fundamental light production or propagation in TLYC, but with only a single sample, we are unable to rule out effects due to non-uniformities in crystal growth or packaging.

\ifCLASSOPTIONcaptionsoff
  \newpage
\fi



%
\bibliographystyle{IEEEtran}
\bibliography{biblio}

\end{document}